# Measuring Technical Debt in AI-Based Competition Platforms

Dionysios Sklavenitis
Hellenic Open University
Patras Greece
sklavenitis.dionysios@ac.eap.gr

Dimitris Kalles
Hellenic Open University
Patras Greece
kalles@eap.gr

## ABSTRACT

Advances in AI have led to new types of technical debt in software engineering projects. AI-based competition platforms face challenges due to rapid prototyping and a lack of adherence to software engineering principles by participants, resulting in technical debt. Additionally, organizers often lack methods to evaluate platform quality, impacting sustainability and maintainability. In this research, we identify and categorize types of technical debt in AI systems through a scoping review. We develop a questionnaire for assessing technical debt in AI competition platforms, categorizing debt into various types, such as algorithm, architectural, code, configuration, data etc. We introduce *Accessibility Debt*, specific to AI competition platforms, highlighting challenges participants face due to inadequate platform usability. Our framework for managing technical debt aims to improve the sustainability and effectiveness of these platforms, providing tools for researchers, organizers, and participants.

## 1 INTRODUCTION

In recent years, progress in Artificial Intelligence (AI) has led to the integration of AI components into an increasing number of software engineering projects, now known as AI-based systems [14, 23]. Even though experienced development teams deliver these projects, they present several new challenges [24].

A most significant challenge is the emergence of new types of technical debt related to AI, which accumulate alongside those already present in traditional software systems [3], negatively impacting their sustainability and maintainability.

Technical Debt in AI systems refers to the long-term maintenance challenges and hidden costs caused by rapid development and deployment. It includes issues like boundary erosion, entanglement, hidden feedback loops, and data dependencies [1], as well as the need for robust abstractions, modular design, and continuous monitoring, which complicate system updates and maintenance over time.

There is a considerable body of recent research on determining, describing, and classifying the types and characteristics of technical debt in AI systems [25], with the primary aim to come up with principles, guidelines, patterns and tests that will ensure the sustainability and maintainability of these systems [30].

A particularly interesting category of such projects are AI-based competition platforms, commercial, academic – research or hybrid, which are utilized by organizations, companies, and universities [21, 25-29] to competitively identify solutions to a problem, while at the same time training participants, either on technical aspects or on soft aspects of AI [16]. Such platforms can vary significantly depending on the scientific field they focus on (medical, robotics, public health, environmental science, ethical AI, computer Vision, NLP, training, etc.), on the community they target (researchers, academics, students, professionals, AI developers and hobbyists), on the infrastructure utilised (open source, closed source, cloud computing), on the specific features they offer (code sharing, custom metrics), and on the cost of using them (free, paid). Monetary prizes, recognition, job opportunities, or academic credits may be offered as incentives [25].

Competitions remain active for a predetermined period, ranging from a few hours to several months [15], during which participants can submit their proposals to solve the given problem.

During the competition, participants emphasize rapid prototyping and iterative design processes. They quickly iterate on their ideas for the problem and experiment with mechanics and game elements to find what works best within the given timeframe. The focus is on creating playable prototypes rather than polished, final products.

After the submission deadline, organizers evaluate the proposed solutions, and the winner is determined [15]. The criteria for rewarding the winner are defined based on the nature of the problem, with the most common being the efficiency of the algorithm (fast response, low resource usage, etc.). Based on frequency and organizational details, competitions can be further classified as One-time, Recurring, Continuous, Series, or Tournaments.

AI-based gaming competitions have gained significant ground in recent years within the academic community [31]. Due to their engaging approach of interacting with students, they accelerate research while simultaneously serving as a premier tool for instruction [15, 16, 33].

Despite their success and student participation, it is observed that within the context of a competition hosted on a university platform, it is common for students to engage only once, primarily driven by the objective to fulfill course requirements. During this process, participants develop algorithms without a clear indication of their adherence to established programming practices or the fundamental principles of software engineering for AI (SE4AI) [3, 4, 32, 34]. This lack of adherence may lead to the gradual accumulation of technical debt within the platform itself. Furthermore, competition organizers currently lack a quantifiable method to evaluate the quality of their platform. This deficiency hampers their ability to prevent the accrual of long-term technical debt, which could undermine the sustainability and maintainability of the platform, thereby diminishing the competition's credibility.

Motivated by the aforementioned objectives, we sought to address the research gap observed in managing technical debt through the application of best practices and fundamental software engineering principles [4], specifically within AI-based competition platforms in the university setting.

This is achieved as follows: for researchers, by identifying new types of technical debt; for organizers and participants, by enriching the available technical debt detection tools; for educators, by offering a straightforward and direct method to train their students in identifying and combating technical debt; and for the student participants, by providing the opportunity to evaluate their proposals before submission.

The objective of this research is to identify the principal factors that contribute to the accrual of technical debt in AI-based systems. This is done through a scoping literature review, then, by classifying all identified factors according to the type of debt they represent. Based on this classification, a questionnaire will be developed to serve as an assessment instrument for organizers and participants of AI competitions to help them assess the extent of technical debt present either on their platforms or within their submissions, respectively.

## 2 RELATED WORK

As developers incorporated AI into applications, they encountered unique challenges, highlighting the need to redefine software engineering. This spurred researchers to investigate AI-specific issues, adapt existing methods, and innovate within the field, leading to the development of new areas such as SE4AI. Equally significant proves to be the contribution of AI-based competition platforms to AI research and education, as well as to the development of soft skills such as collaboration, communication and soft aspects such as ethics and human-centered design.

### 2.1 Technical Debt in AI-based Systems

Sculley et al. [1] introduced the concept of "Hidden Technical Debt" in machine learning (ML) systems, highlighting the unique challenges of managing such debts. Subsequent research by Liu et al. [2] identified extensive technical debt in leading deep learning frameworks, particularly within design, requirements, and algorithms. A. Stewart [3] and Recupito et al. [4] further explored ML-specific technical debts and introduced Artificial Intelligence Technical Debt (AITD) to address gaps in managing code and architectural debts.

Code smells are a primary factor contributing to technical debt. Gesi et al. [5] in 2022 identified five new smells affecting the maintainability of deep learning systems, while Zhang, Cruz, and Deursen [6] uncovered 22 distinct smells across different pipeline stages. Additionally, Foidl, Felderer, and Ramler [7] explored data smells, providing detection methods.

Significant strides in understanding technical debt in AI systems were made by Bogner et al. [8], who identified types such as data and model debts and described 72 antipatterns with 46 solutions. H. Washizaki et al. [9, 35] applied software engineering design patterns to enhance ML system architecture and efficiency, and Heiland, Hauser, and Bogner [10] categorized 70 design patterns for various applications.

Nascimento et al. [11] and Serban et al. [12] reviewed software engineering practices tailored for ML systems, emphasizing the need for methodologies specific to ML. V. Lenarduzzi et al. [13] and Martínez-Fernández et al. [14] highlighted the evolving nature of software quality metrics and synthesized SE practices, examining numerous studies to better integrate software engineering in AI development.

### 2.2 AI-based Competition Platforms

AI competition platforms are key in advancing software engineering methodologies and provide essential training in AI-related soft skills. Togelius [15] discusses the organization of successful AI game competitions, emphasizing their role in spurring innovation and rapid development, albeit sometimes at the expense of long-term sustainability. D. Kalles [16] explores how integrating AI and software engineering in education can mitigate document debt and develops soft skills using platforms like RLGame. The RoboCup, as analyzed by Genter, Laue, and Stone [17], exemplifies how such competitions train robotics and AI students, enhancing teamwork skills in AI systems without pre-coordination.

Competitions like Project Malmo [18], built on Minecraft, highlight how AI challenges can advance rigorous research and develop skills in strategic planning and real-time collaboration. Similarly, Giagtzoglou and Kalles [19] and Salta, Prada, and Melo [20] demonstrate how gaming ecosystems like RLGame and Geometry Friends contribute to AI education and research. Konen [21] introduces the General Board Game (GBG) framework, promoting its use in educating diverse AI agents across various games. Pavao et al. [22] propose using CodaLab in competitive AI settings to manage technical debt, thus bolstering the training of AI professionals and ensuring the sustainability of AI systems in dynamic environments.

# 3 RESEARCH METHODOLOGY & FINDINGS

## 3.1 Goal and research questions

The integration of artificial intelligence components into software engineering systems has catalyzed the formation of specialized research domains such as SE4AI and SE4ML [35]. These domains dedicate their focus to applying the principles from one specialized area to another, which in turn, expands the range of sources and causes of technical debt associated with these advancements. This expansion necessitates the formulation of novel categories of technical debt that encapsulate these new factors and causes, thereby enhancing their classification and analysis. Motivated by this need, we opted for a scoping review as our research methodology. This approach enables a thorough examination of the varied facets by systematically mapping out the current body of literature. The organization of our study involved multiple stages, the specifics of which are detailed in the Appendix A.

The objectives of our study were defined as follows:

1. To systematically identify, classify, and categorize the factors contributing to technical debt in AI-based systems, drawing upon extant scholarly works.
2. To elucidate each type of technical debt by providing a succinct definition, a brief description of its impacts and an illustrative example of its application
3. To ascertain the types of technical debt prevalent in current AI-based competition platforms.
4. To develop a custom questionnaire for each identified type of technical debt, aimed at quantifying its presence in AI-based competitive environments.
5. To propose methodologies for the quantification of technical debt and to explore the potential benefits of addressing it in the context of an AI-based competition platform.

Based on our goal, we defined the following research questions (RQs):

RQ1: What are the most significant types of technical debt recorded in AI-based systems?

RQ2: How can we measure the technical debt of an AI – based competition platform?

## 3.2 Findings

From the research and study of the articles we identified, we found that the management of technical debt is associated with several factors. Therefore, to answer RQ1, we categorized these factors into 18 distinct types. For each type, we provided a definition, a brief description of its implications, and a hypothetical scenario tailored for AI competition platforms, to facilitate study and interpretation. An initial classification of studies by type of technical debt and year of publication identified is presented in Table 1. Subsequently, we utilized the categorization from RQ1 to develop the measurement approach for technical debt, as required by RQ2. Thus, we defined the types shown in Table 1.

**Table 1. Types of Technical Debt in AI-based Systems**

| S/N | Type | Number of Documents | Publication Span Period |
|---|---|---|---|
| 1 | Algorithm | 4 | 2020 - 2023 |
| 2 | Architectural | 10 | 2016 - 2023 |
| 3 | Build | 6 | 2012 - 2022 |
| 4 | Code | 7 | 2019 - 2022 |
| 5 | Configuration | 4 | 2015 - 2021 |
| 6 | Data | 9 | 2017 - 2022 |
| 7 | Defect | 5 | 2017 - 2021 |
| 8 | Design | 4 | 2014 - 2021 |
| 9 | Documentation | 4 | 2020 - 2022 |
| 10 | Ethics | 5 | 2018 - 2022 |
| 11 | Infrastructure | 3 | 2019 - 2021 |
| 12 | Model | 13 | 2015 - 2022 |
| 13 | People | 3 | 2017 - 2022 |
| 14 | Process | 2 | 2014 - 2021 |
| 15 | Requirements | 5 | 2019 - 2023 |
| 16 | Self-Admitted (SATD) | 6 | 2016 - 2023 |
| 17 | Test | 4 | 2017 - 2023 |
| 18 | Versioning | 6 | 2017 – 2021 |
| | Total | 100 | |

In this section, we present the nine most significant types of technical debt in the context of AI-based competition platforms.

We identified eight out of the nineteen proposed categories as the most significant, based on their prevalence in the literature [8, 14]. The integration of artificial intelligence components into software has compelled the scientific community to examine new aspects of technical debt. These aspects include Data, Model, Configuration, and Ethics of AI-based systems, as well as modifications in existing technical debt factors of traditional software, such as Architecture, Testing, and Infrastructure [8]. Moreover, the examination of technical debt factors related to the Algorithms used in Deep Learning systems has become essential [2]. Additionally, diverging from our primary criterion, we deemed the examination of Accessibility Debt essential due to its significant impact on participant engagement and the overall success of AI-based competition platforms.

From the organizers' perspective, focusing on these types of technical debt helps ensure that the technical environment of the competition provides a straightforward and efficient method of participation and is optimized for the development and advancement of artificial intelligence. This optimization enhances both the experience and the performance of the participants' models. Additionally, these priorities help maintain a fair and competitive atmosphere, which is vital for the success of an AI-based competition. For participants, these types reflect the direct impact on their ability to compete seamlessly, promptly, effectively and innovate within the given platform. This ensures that their solutions are not only technically correct but also ethically and practically robust.

The remaining types of technical debt identified, are presented in the Appendix B.

### *3.2.1 Algorithm debt*

*Definition:* Algorithm debt refers to technical debt arising from the choice, implementation, and maintenance of algorithms within AI systems [2], excluding issues directly related to the model's structure or data handling. This debt encompasses challenges associated with selecting appropriate algorithms, optimizing their performance, and ensuring they remain suitable as the system scales or as requirements evolves.

*Problem:* The problem of algorithm debt in AI-based systems often stems from the use of algorithms that are either too simplistic or overly complex for the task at hand, poorly optimized, or inefficient in terms of computational resources. Such choices may initially simplify development but can lead to increased costs and reduced system performance in the long run. For instance, an algorithm that is not scalable might handle initial data volumes well but becomes a bottleneck as data grows, requiring costly re-engineering efforts.

*Example:* In the context of AI-based competition platforms, an example of algorithm debt could occur when an algorithm designed for matchmaking in online games does not effectively adapt to changes in player skills and behaviors over time. Initially, the algorithm may work well, but as the variety of players increases, its inability to adapt could result in poor matchmaking, increased wait times, and player dissatisfaction. The platform may then face significant technical debt as the original algorithm requires substantial modification or replacement to meet the evolved needs of its user base.

This example highlights how crucial it is to anticipate the long-term needs of the system when choosing algorithms, and to plan for their evolution as part of the platform's ongoing development strategy to avoid accruing algorithm debt. This example illustrates how the need for rapid development in competitive AI platforms can lead to significant algorithm debt, impacting the platform's long-term capability to perform reliably and accurately.

*Stakeholder:* Participants must address any potential glitches or incompatibilities in the frameworks or algorithms they use. This is crucial to prevent issues that could impair the performance of their AI solutions, such as slower processing times or errors during execution.

### *3.2.2 Architectural Debt*

*Definition*: Architectural debt refers to the complexities arising from the intricate integration of architectural components with their foundational data within AI-based systems [8].

*Problem*: This type of debt can precipitate intricate and indeterminate dependencies between system components and their corresponding datasets. It often results in architectures that are challenging to evaluate due to their complex compositions. Moreover, it can lead to the emergence of undeclared entities that utilize AI models, further complicating the governance and maintenance of these systems.

*Example*: At a global AI-based game development competition, architectural debt in gaming platforms posed significant challenges. The systems were designed to adapt dynamically to user behaviors using complex architectures that integrated multiple data sources and AI components. This complexity led to unpredictable game behaviors, where player input changes unexpectedly altered AI responses due to hidden architectural dependencies. Additionally, undocumented consumption of extra data by some components caused inconsistencies and performance issues, complicating the assessment and modification of games by developers. This severely impacted the competition experience and the performance of AI models, highlighting how architectural debt can undermine functionality and innovation in AI-driven competitions.

*Stakeholder*: Organizers. Ensuring a well – structured system architecture allows organizers to maintain and scale the platform efficiently. This involves good practices in separating concerns, managing dependencies, and avoiding tightly coupled components, which collectively ensure that the system remains flexible and manageable as it evolves.

### *3.2.3 Configuration Debt*

*Definition*: Configuration debt encapsulates the deficiencies prevalent within the configuration frameworks of AI-based systems. This term specifically refers to the complexities and shortcomings of configuration processes, including the use of convoluted, insufficiently documented, unversioned, or untested configuration files [1, 3, 8].

*Problem*: Configuration debt introduces significant vulnerabilities to machine learning systems by fostering errors in configuration that can deplete valuable time and computational resources and cause delays in production. This debt hampers the ability to accurately modify and understand configurations, complicating the evaluation of the effects of changes and the comparison of configurations across different iterations. Moreover, poor management of configuration settings intensifies these issues by obstructing the precise specification, tracking, and reproducibility of configuration alterations. This results in difficulties in replicating experiments and ensuring system reliability. Effective management of configuration settings is crucial to alleviate these problems and enhance the efficiency, reproducibility, maintainability, and transparency of AI-based systems.

*Example*: During a high-profile AI-based game competition, one team struggled with significant configuration debt. Their gaming AI had multiple configuration files that were complex and poorly documented, making it difficult for new team members to understand and modify the settings efficiently. As the competition progressed, the need to adapt to opponents' strategies became critical. However, due to unversioned and untested configurations, changes made under pressure led to errors that were not detected until too late. This resulted in the AI performing unpredictably, costing the team crucial matches and demonstrating the impact of configuration debt on competitiveness and system reliability.

Effective configuration management allows rapid iteration and model improvement, which are essential in a competitive AI environment. Organizers need to ensure that configurations can be easily managed and modified to accommodate diverse models and strategies that participants might employ.

*Stakeholder*: Organizers: Effective configuration management allows rapid iteration and model improvement, which are essential in a competitive AI environment. Organizers need to ensure that configurations can be easily managed and modified to accommodate diverse models and strategies that participants might employ.

*3.2.4 Data Debt*

*Definition*: Data debt pertains to the shortcomings in the collection, management, and application of data within AI-based systems [8], characterized by issues such as poor data quality, unregulated data dependencies, and inadequate data relevance [7].

*Problem*: These deficiencies can compromise the efficiency and precision of AI models, posing challenges to the system's reliability and future adaptability. Data debt introduces risks that may impede the ongoing development and operational effectiveness of AI systems, potentially leading to erroneous outputs and strategic misalignments in long-term system evolution.

*Example*: In a high-profile AI-based gaming competition, a team faced setbacks from data debt. Their AI model, designed to dynamically adapt strategies, used outdated and poorly curated datasets, compromising data relevance and quality. This affected the model's decision-making, often resulting in ineffective strategies and failure to anticipate opponent moves. Unmanaged data dependencies, only discovered during the competition, highlighted the need for real-time data processing and underscored the importance of rigorous data management to maintain AI effectiveness and system scalability in dynamic settings.

Data quality is paramount in AI competitions. Organizers must ensure that the data used is accurate, complete, and relevant. Managing data debt effectively prevents issues such as overfitting or poor generalization, which can severely affect the outcomes of the competition.

*Stakeholder:* Organizers: Data quality is paramount in AI competitions. Organizers must ensure that the data used is accurate, complete, and relevant. Managing data debt effectively prevents issues such as overfitting or poor generalization, which can severely affect the outcomes of the competition.

Participants: Quality and integrity of data are paramount in training effective and reliable AI models. Participants must ensure their data is accurate, complete, relevant, and trustworthy to avoid issues like overfitting, underfitting, or biased results, which could severely affect their model's performance.

*3.2.5 Model Debt*

*Definition*: Model debt refers to the accumulation of suboptimal practices within the lifecycle of artificial intelligence models, encompassing the design, training, and management phases. This encompasses issues related to inadequate feature selection, improper tuning of hyperparameters, and ineffective strategies for model deployment [1, 2, 8].

*Problem*: Model debt manifests through several challenges, such as poorly chosen features, neglected adjustments of hyperparameters, and deficient deployment architectures. Such deficits may complicate the maintenance of models and diminish their predictive accuracy, thereby undermining the system's overall effectiveness and reliability.

*Example*: In an AI-based game competition, one team faced significant model debt issues. Their AI model, initially promising, suffered from too narrow feature selection focused on short-term gains and lacked strategic depth. Additionally, the hyperparameters were not tuned to the game's dynamic nature, and the deployment strategy failed to adapt to evolving scenarios. These shortcomings led to the model underperforming in critical matches, unable to accurately predict or counter the diverse strategies of competitors.

This is critical because it directly affects the quality and performance of the AI models used in the competition. Ensuring that models are well-validated, free from bias, and perform robustly across different scenarios is key to maintaining the integrity and fairness of the competition.

*Stakeholder:* Organizers: This is critical because it directly affects the quality and performance of the AI models used in the competition. Ensuring that models are well-validated, free from bias, and perform robustly across different scenarios is key to maintaining the integrity and fairness of the competition.

Participants: Participants need to ensure that their models are robust and validated correctly before serving. This includes avoiding feedback loops and ensuring that the model can be debugged and tested comprehensively. This type of debt is vital to prevent deploying models that might perform well in training but fail in practical scenarios.

*3.2.6 Ethics Debt*

*Definition:* Ethics debt refers to the shortcomings concerning the ethical dimensions of AI-based systems, including issues such as algorithmic fairness, prevalent prediction biases, and a lack of transparency and accountability [8].

*Problem:* The realm of AI ethics encounters multiple challenges that stem from a diminished influence over decision-making processes, insufficient enforcement mechanisms, and an over-reliance on ethical guidelines rather than binding legal standards.

This often results in the neglect of broader social contexts and relationships, contributing to a lack of diversity within the AI community and the exclusion of vital ethical considerations. The ramifications of ethics debt are manifold, encompassing unintended harmful effects and the malicious use of AI technologies. Such consequences include job displacement, unsupervised experimental applications, and significant data breaches. Moreover, there exists a substantial risk of developing biased algorithms and unsafe products, precipitously deploying immature applications, and the potential exploitation of AI technologies by malicious entities, such as criminal hackers.

*Example:* In an AI-based game competition, a team deployed an AI model designed to predict and counter opponents' moves. Unfortunately, the model exhibited significant ethics debt due to overlooked issues like algorithmic fairness and transparency. The model, trained predominantly on data from games played by a non-diverse group, failed to fairly assess strategies used by a wider range of competitors. This bias led to inaccurate predictions and strategic blunders, compromising the team's performance. The incident highlighted the consequences of neglecting ethical considerations in AI development, demonstrating how ethics debt can undermine fairness and effectiveness in competitive environments.

*Stakeholder:* Participant: Understanding and adhering to ethical guidelines set by the competition is essential. This includes knowing the rules, how to raise questions, and how to handle potentially conflicting interpretations of the competition's terms. Ethical considerations are crucial to ensure that the competition is fair and that all participants operate on a level playing field.

*3.2.7 Infrastructure Debt*

*Definition:* Infrastructure debt encapsulates the inadequacies inherent in the implementation and operational management of artificial intelligence (AI) pipelines and models [1, 8, 11, 30].

*Problem:* This type of debt introduces significant operational and reproducibility challenges within AI systems. It frequently manifests as overly complex infrastructures that integrate multiple AI pipelines, leading to suboptimal resource distribution for the training and testing of AI models. Additionally, it exacerbates the difficulty in effectively monitoring and debugging AI systems, thereby compromising their reliability and performance.

*Example:* During a high-profile AI-based game competition, infrastructure debt became evident as teams struggled with the systems set up for training and testing their game-playing models. The competition's infrastructure was initially designed to handle multiple concurrent AI pipelines efficiently. However, as the competition progressed, it became clear that there were significant deficiencies in resource allocation and system integration. Some teams experienced delays in model training due to limited GPU resources, while others faced challenges in consistently reproducing game strategies due to varying system performances. Additionally, the lack of robust monitoring and debugging tools meant that identifying and resolving these issues was time-consuming, leading to uneven playing fields and questioning the fairness and integrity of the competition. This scenario highlights how infrastructure debt can severely impact the operational success of AI-driven initiatives in competitive environments.

*Stakeholder:* Organizer: Reliable and scalable infrastructure is crucial for handling the computational demands of AI models, especially when multiple participants are involved. Organizers must manage infrastructure debt by ensuring the system is robust, scalable, and capable of supporting the high concurrency levels typically seen in competitions.

*3.2.8 Test Debt*

*Definition:* Test debt encompasses the shortcomings in testing practices within AI systems, particularly highlighting the insufficient testing of AI models and pipelines, as well as a deficiency in advanced testing methodologies necessary for evaluating data quality and distribution [8, 30] .

*Problem:* The probabilistic characteristics of certain AI algorithms add a layer of complexity to the testing processes. These complexities make it especially challenging to ensure comprehensive testing coverage and reliability, thereby increasing the risk of undetected issues in AI system behavior under varied operational conditions. This type of debt may lead to less predictable and potentially unreliable AI system performance.

*Example:* In an AI-based game competition, various AI models were pitted against each other in a strategy game. However, due to existing test debt, the competition faced challenges. The AIs had not been adequately tested for their ability to handle the stochastic elements of the game, such as random events and unpredictable opponent strategies. This lack of rigorous testing resulted in some AI models performing inconsistently, displaying erratic behavior, or failing to adapt to new scenarios presented during the competition. The event highlighted critical gaps in testing practices, emphasizing the need for more robust testing frameworks to ensure AI reliability and effectiveness in dynamic environments.

*Stakeholder:* Participant: Ensuring that their model and its components, such as hyperparameters and training environments, are thoroughly tested is crucial. This includes validating the reproducibility of results and the stability of the model in the competition's game environment, which is essential for demonstrating effectiveness and reliability.

In the context of AI/ML competition platforms, a critical factor for their successful operation is the ease and speed with which potential participants can utilize the platform's components to engage with the assigned tasks. Consequently, we have defined a new type of technical debt, specific to AI-Based Competition Platforms, as *Accessibility Debt*. This definition addresses the challenges associated with the ease and speed of immediate use of the platform technologies, which, if not adequately addressed, discourage participants, thereby devaluing the competitions.

*3.2.9 Accessibility Debt*

*Definition:* Accessibility Debt refers to the barriers that participants encounter due to the lack of immediate usability of platform technologies.

*Problem:* The primary issue with Accessibility Debt is that it can dissuade potential participants from engaging with AI/ML competition platforms. This is due to the challenges they face in utilizing the platform's components quickly and efficiently. If these barriers are not addressed, the competitive value and attractiveness of such platforms diminish, leading to decreased participation and devalued competitions.

*Example:* Consider an AI-based competition platform where participants must navigate a complex setup process before they can begin working on their assigned tasks. If the process involves multiple steps, unclear instructions, or requires significant troubleshooting, participants may become frustrated and disengaged. This scenario exemplifies Accessibility Debt, as the immediate usability of the platform is compromised, hindering participant involvement and satisfaction.

*Stakeholder:* Organizer: They are responsible for ensuring that the platform is accessible and user-friendly, providing necessary tools and resources to participants efficiently. By addressing Accessibility Debt, organizers can enhance participant engagement and the overall success of the competitions.

## 4 PROPOSED METHOD FOR MEASURING TECHNICAL DEBT

Technical debt is a multifaceted concept that affects all stages of the software development lifecycle. The unique challenge presented by AI-based systems, to be "just like" traditional software until they're not [13], exacerbates the difficulty in identifying and addressing technical debt. This issue is especially pronounced on platforms for AI/ML competitions, where applications might only be deployed once. It is crucial that participants are informed about the importance of technical debt and are able to identify its manifestations. Such awareness is fundamental for their training in best practices and equips them sufficiently to excel as proficient practitioners in their future professional endeavors.

In an effort to aid both participants and organizers in recognizing and quantifying technical debt, we devised a questionnaire detailed in the Appendix C, which categorizes technical debt according to the types identified in RQ1, as well as the new proposed type and answers to RQ2. To our knowledge, there are no comparable initiatives. The endeavor documented in [30] focuses solely on ML applications, with an emphasis primarily on testing and monitoring. Although we incorporated some of these questions, our questionnaire has been expanded to encompass all debt types that surfaced in RQ1 and [8].

The questionnaire employs a rating scale ranging from 1 (minimal) to 5 (maximum), though these ratings are not disclosed to users, along with the response options 'Not Applicable' and 'Don't Know/Don't Answer':

1. This allows participants to quantitatively assess the level of technical debt in their proposed implementations while simultaneously developing soft skills [16, 31] and familiarizing themselves with the nuanced aspects [19] - which are increasingly critical - of AI applications.
2. This enables organizers to evaluate their platforms in order to prevent the accrual of long-term technical debt resulting from suboptimal implementation decisions (poor SE4AI choices), thereby enhancing the viability and maintainability of their platforms.

In the questionnaire designed to quantify technical debt, each response plays a specific role in the scoring system that ultimately reflects the level of technical debt associated with a given proposal or platform. A "YES" response, which signifies adherence to practices known to mitigate technical debt, results in a negative score, thereby indicating a reduction in technical debt. Conversely, a "NO" response suggests a deviation from these practices, and is consequently assigned a positive score, indicating an increase in technical debt. Responses marked as "Not Applicable" receive a score of 0, reflecting their neutrality in terms of impacting the technical debt calculation. Similarly, responses of "I Don't know/I Don't answer" are treated the same as "NO" responses, as they both imply the presence or potential accumulation of technical debt. The rationale for this scoring method is founded on the assumption that affirmative answers demonstrate compliance with best practices that lower technical debt, whereas negative or uncertain responses indicate areas where technical debt could either be present or likely to accumulate. The cumulative score derived from all responses thus provides a quantitative measure of the total technical debt associated with the entity being evaluated.

In our approach to quantifying technical debt, we delineate the following components for each type of debt assessed:

1. **Question**: This represents the specific query posed to stakeholders, aimed at gauging their practices related to technical debt.
2. **Stakeholder**: This identifies the individual or group responsible for responding to the question, thereby ensuring that the query is directed to those with the most relevant experience or decision-making authority.
3. **Score**: Assigned on a scale from 0 to 5, this quantifies the response, with different scores reflecting varying degrees of adherence to best practices in technical debt management.
4. **Justification**: This provides the rationale behind the necessity of posing the question, linking it to its potential impact on technical debt.
5. **Example**: A typical scenario or case is presented to clarify the question and assist the stakeholder in understanding and responding appropriately.

An aggregated presentation of the questions categorized by type of technical debt is included in Table 2 of our document. Due to the extensive range of questions developed for this assessment, we highlight five representative questions in the main text. The complete set of questions, however, is detailed in the Appendix C for interested readers seeking comprehensive insights into the questionnaire framework and its applications.

**Table 2. Number of Questions by Technical Debt Type**

| Type | Number of Questions |
|---|---|
| Accessibility | 3 |
| Algorithm | 1 |
| Architectural – Design | 5 |
| Build | 2 |
| Code | 1 |
| Configuration | 3 |
| Data | 6 |
| Defect | 5 |
| Documentation | 5 |
| Ethics | 2 |
| Infrastructure | 4 |
| Model | 3 |
| People – Social | 2 |
| Process | 3 |
| Requirements | 5 |
| Self-Admitted (SATD) | 10 |
| Test | 7 |
| Versioning | 3 |
| Sum | 68 |

## 5 DISCUSSION

By exploring the multidimensional nature of technical debt in AI-based competition platforms, we have highlighted a multitude of debt types that are important to both organizers and participants. These debts highlight the complexity of managing AI systems and span the following aspects: algorithms, architecture, configuration, data, model, accessibility, ethics, infrastructure, and testing. Notably, algorithm and architecture debts profoundly affect system performance and scalability. For example, algorithm debt, resulting from the selection and maintenance of suboptimal algorithm libraries, can severely hinder the adaptability of AI systems. Similarly, architectural debt, with its complex dependencies and integration challenges, hinders system evaluation and flexibility, thereby degrading user experience and system robustness. Ethical debt can also undermine competitive success when competitions prioritize technical robustness exclusively, disregarding other crucial factors. For example, competitions which fail to provide reliable test and training data eventually, compromise fairness and trust. Additionally, insufficiently training for participants to adhere to legal standards [36], which complement ethical guidelines, when the domain so requires, and inadequate instruction on implementation rules or conflict resolution, can further jeopardize clarity and fairness.

For organizers, understanding and prioritizing these types of technical debt is crucial in maintaining a fair and competitive atmosphere essential for the success of AI-based competitions. This involves a meticulous focus on not just the immediate functionality but also the long-term sustainability of the competition platform. Organizers must adopt proactive strategies to mitigate these debts, such as refining system architecture, ensuring robust configuration management, and fostering ethical practices to support diverse AI models and strategies efficiently.

Participants, on the other hand, face direct consequences of these technical debts, influencing their capacity to innovate and compete effectively. Addressing algorithm, configuration and ethical debts, for instance, is crucial for participants to ensure their AI solutions are not only technically adept but also fair, transparent and adaptable to the evolving competitive landscape.

### 5.1 Accessibility Debt: A New Liability

Our study introduces a novel category – Accessibility Debt, specific to AI-Based Competition Platforms – which highlights the barriers participants encounter due to the lack of immediate usability of platform technologies. This form of debt is particularly detrimental as it can dissuade potential participants, thus diminishing the competitive value and attractiveness of AI-based platforms. Organizers need to address this by simplifying access to necessary tools and resources, ensuring that participants can engage with the platform effectively and without undue delays.

### 5.2 A Questionnaire as a Quantifiable Assessment Tool

The proposed questionnaire, derived from an initial pool of about 160 questions from our research, comprehensively covers the types of technical debt predominant in AI-based systems, as outlined in RQ1. These questions, grouped by type of technical debt, assess stakeholder actions from the highest (abstract) level, such as AI system architecture and design, to the lowest (detailed) level, like code writing and system configuration, measuring their impact on reducing technical debt. Given the potential challenges posed by the large number of questions, we prioritized them by selecting those with a significance weight of 3 and above, while keeping the score hidden. This approach reduces users' cognitive load with a simpler, more substantial questionnaire, avoiding confusion while focusing on the most critical checks.

To enhance the reader's understanding of the questionnaire's application, we will provide an example for each stakeholder, using two of the nine technical debt categories listed in the article.

**Table 3. Organizer Accessibility Debt Example**

| Question | Score | Answer | Calculated Score |
|---|---|---|---|
| Have you conducted usability testing to identify and address potential barriers in the platform setup process? | 5 | YES | -5 |
| Have you integrated adaptive user interfaces that personalize the setup experience based on participants' skill levels | 4 | NO | 4 |
| and preferences? | | | |
| Have you implemented feedback mechanisms for participants to report accessibility issues and suggest improvements? | 3 | I Don't Know/I Don't Answer | 3 |
| **Overall Rating** | | | **2** |

**Table 4. Participant Model Debt Example**

| Question | Score | Answer | Calculated Score |
|---|---|---|---|
| Are you detecting direct feedback loops or hidden feedback loops? | 4 | NO | 4 |
| Is model quality validated before serving? | 5 | YES | -5 |
| Does the model allow debugging by observing the step-by-step computation of training or inference on a single example? | 3 | I Don't Know/I Don't Answer | 3 |
| **Overall Rating** | | | **2** |

According to the stakeholder responses depicted in the examples from Tables 3 and 4, the platform assessment for the specified types reveals a quantifiable technical debt of 2 units each. Each interested party will independently replicate the example process for each question (Organizers: 41, Participants: 41) in Appendix C relevant to their context. The cumulative results from both Organizers and Participants will indicate the degree (percentage) of technical debt identified on the platform. A total final sum less than or equal to 0 signifies that the platform adheres to SE4AI best practices, indicating zero debt. Conversely, a positive sum indicates a high level of technical debt on the platform.

## 6 THREATS TO VALIDITY

In all phases of our research we tried to mitigate potential threats to validity that are very common in this type of research. However, our research is subject to limitations.

**Internal validity**. To reduce potential bias due to the subjective interpretation of researchers, we defined during our research (RQ1), a comprehensive set of keywords related to technical debt in AI-based systems and competing platforms. We selected five datasources for our search to broaden the results by sticking mainly to peer-reviewed papers. We also performed snowballing to ensure inclusion of papers that may have been missed in the original search.

In the first selection phase we studied the title, abstract and keywords of the results obtained. Each article was read by the first author in its entirety in detail and comments were made regarding content, key points and findings relevant to our research.

The second author reviewed these comments, assessed the quality of the findings while the authors discussed any papers in doubt and made a joint decision thus ensuring that the papers were suitable for inclusion in the final results. These studies were selected based on the criteria they set, while a similar procedure was followed for data extraction and analysis.

Despite faithful adherence to the protocol, due to subjectivity there is a possibility that other researchers who wish to use our strategy may come up with slightly different results.

**External validity**. The area of technical debt in AI-based systems is an area that is constantly evolving and changing dynamically either by modifying existing ones or by including new categories. Several times, issues that have already been highlighted are re-examined under a different perspective or different framework of application, highlighting new dimensions.

As a consequence, nuanced findings may emerge on the definition, scope, and impact of technical debt in AI-based systems as they mature and are increasingly adopted in software engineering. Especially in an unexplored area like that of AI-based competition platforms, we think that's to be expected.

## 7 CONCLUSION & FUTURE RESEARCH

In this study, we collected, classified, and analyzed the dominant types of technical debt for AI-based systems. Utilizing these classifications, we developed a questionnaire as a self-assessment tool for measuring technical debt on AI-based competition platforms. Our aim is to bridge the observed measurement gap of technical debt on such platforms.

We have discovered that managing technical debt in AI-driven platforms necessitates a comprehensive strategy tailored to every unique phase of the platform's lifecycle. The intricacy of these systems arises not only from the incorporation of AI elements but also from the enhanced interplay with human factors, exerting a more significant influence than observed in conventional software engineering contexts.

We have identified a new type of technical debt, *accessibility debt*, which highlights the steepness of the learning curve for some competition platforms, thus diminishing their competitive value and attractiveness. Our findings resonate well with and extend the current literature on technical debt in software engineering, by contextualizing these debts within the unique demands of AI competition platforms.

As the next step in our research, we plan to test the effectiveness of the questionnaire in practice by having it evaluated: a) by organizers (professors) who teach AI software engineering and use competition platforms to educate their students, b) by participants (undergraduate students of computer science departments), who are taught AI software engineering and use competition platforms to develop their proposals. We expect to empirically determine: a) the percentage of technical debt that can be detected on a competition platform (by organizers and participants), b) the extent to which it can train students to identify technical debt, c) the degree to which it contributes to reducing the long-term debt of the platform itself.

Additionally, as the creators of the questionnaire, we will have the opportunity to a) assess whether the nine most significant types of technical debt identified in this article are equally important across various competition platforms and among diverse stakeholders who will use the questionnaire, b) observe variations in the use and outcomes of the questionnaire under two conditions: i) when the characteristics of the stakeholders are known in advance but control variables have not been defined, ii) when the characteristics of the stakeholders are not known in advance, but appropriate control variables, such as education, years of coding experience, frequency of participation in competition platforms, dedicated training time, have been defined. Furthermore, we will endeavor to determine to what extent it is possible to assess similar systems in commercial environments, including both proprietary and closed-source platforms such as Azure and Kaggle.

# APPENDIX A

## A.1 Research Objective

The objective of this scoping review is to systematically map the literature on technical debt in Artificial Intelligence based systems, identifying the main contributors, methodologies used for evaluation, and existing research gaps. This will help formulate strategies to mitigate the technical debt and guide future studies in this rapidly evolving field.

## A.2 Goal and research questions

The integration of artificial intelligence components into software engineering systems has catalyzed the formation of specialized research domains such as SE4AI and SE4ML. These domains dedicate their focus to applying the principles from one specialized area to another, which in turn, expands the range of sources and causes of technical debt associated with these advancements. This expansion necessitates the formulation of novel categories of technical debt that encapsulate these new factors and causes, thereby enhancing their classification and analysis. Motivated by this need, we opted for a scoping review as our research methodology. This approach enables a thorough examination of the varied facets by systematically mapping out the current body of literature.

The objectives of our study were defined as follows:

6. To systematically identify, classify, and categorize the factors contributing to technical debt in AI-based systems, drawing upon extant scholarly works.
7. To elucidate each type of technical debt by providing a succinct definition, a brief description of its impacts and an illustrative example of its application

Based on our goal, we defined the following research questions (RQs):

RQ1: What are the most significant types of technical debt recorded in AI-based systems?

## A.3 Search Strategy

The strategy followed includes the bibliographic sources (databases, websites), the search terms, the search strings, the definition of the inclusion and exclusion criteria and the selection process followed for their inclusion.

## A.4 Data Source Selection

In order to record a comprehensive set of relevant bibliography, the search engines Google Scholar and the electronic databases ACM Digital Library, IEEE Xplore, Scopus and Springer were searched.

## A.5 Search Terms

As can be understood, the causes that create and the factors that influence technical debt are numerous and diverse and therefore, it was deemed necessary to investigate its more general correlation with Artificial Intelligence/Machine Learning applications. Thus, we selected the terms below as most representative of completeness in generating the search queries.

**Terms**: Artificial Intelligence, AI, Machine Learning, ML, Technical Debt, Software Engineering, AI-Based Systems.

## A.6 Search Strings

To define the search queries, we used appropriate literal and semantic synonyms to include as many results as possible. The following list shows the search queries per data source we searched for:

**Data source**: Google Scholar
**Date limitation**: since 2012 to Feb 2024
**Search String**: ("Technical Debt" AND ("Artificial Intelligence" OR "AI" OR "Machine Learning" OR "ML") AND "Software Engineering") OR ("Technical Debt" AND "AI-Based Systems")
**Results**: 214

**Data source**: ACM:
**Date limitation**: since 2012 to Feb 2024
**Search String**: "Technical Debt" AND ("Artificial Intelligence" OR "AI" OR "Machine Learning" OR "ML") AND ("Software Engineering" OR "SE") OR ("Technical Debt" AND "AI-Based System*")
**Results**: 413
**Data source**: IEEE Xplore
**Date limitation**: since 2012 to Feb 2024

**Search String**: ("Technical Debt" AND ("Artificial Intelligence" OR "AI" OR "Machine Learning" OR "ML") AND ("Software Engineering" OR "SE")) OR ("Technical Debt" AND "AI-Based Systems")
**Results**: 49

**Data source**: Scopus
**Date limitation**: since 2012 to 2024
**Search String**: TITLE-ABS-KEY ("Technical Debt" AND ("Artificial Intelligence" OR "AI" OR "Machine Learning" OR "ML") AND ("Software Engineering" OR "SE")) OR TITLE-ABS-KEY ("Technical Debt" AND "AI-Based Systems")
**Results**: 46

**Data source**: Springer
**Date limitation**: since 2012 to 2024
**Search String**: ("Technical Debt" AND ("Artificial Intelligence" OR "AI" OR "Machine Learning" OR "ML") AND ("Software Engineering" OR "SE")) OR ("Technical Debt" AND "AI-Based Systems")
**Results**: 148

## A.7 Search Criteria

Studies were selected based on their focus on technical debt specifically within AI-based systems.
Inclusion criteria:
I1: Articles from 2012 to February 2024
I2: Articles about AI-based Systems
I3: Articles about AI/ML- based Competitions
I3: Studies included peer-reviewed articles, conference papers that directly addressed the management, implications, or mitigation of technical debt in AI
Exclusion criteria:
E1: Articles not in English
E2: Articles earlier than 2012
E3: Articles or Conference papers not peer-reviewed
E4: Studies not specifically addressing AI Systems
E5: Outdated technologies no longer applicable to current AI paradigms

## A.8 Study Selection

The study selection process was carried out in two phases: initial screening of titles and abstracts for relevance followed by a full-text review to confirm eligibility based on the predefined criteria.
After completing the initial search we reviewed the reference lists of the identified studies and performed forward and backward snowballing to ensure the identification of additional studies that were cited by these articles but may not have been captured in the initial database searches. Also, articles that did not result from the above procedure were manually added. In this phase, one researcher also commented on the studies and highlighted their key points. In the second phase a two-stage screening was conducted independently by two researchers to minimize bias, with discrepancies resolved through discussion or consultation until consensus was reached.

## A.9 Results

A total of 100 studies were selected based on relevance to the topic. These studies provided insights into various aspects of technical debt in AI-based systems. In table 1 we present the distribution of studies by type of debt while in table 2 we present the findings of our research.

**Table 5. Key Types of Technical Debt Identified in AI-based Systems**

| S/N | Type | Number of Documents |
|---|---|---|
| 1 | Algorithm | 4 |
| 2 | Architectural | 10 |
| 3 | Build | 6 |
| 4 | Code | 7 |
| 5 | Configuration | 4 |
| 6 | Data | 9 |
| 7 | Defect | 5 |
| 8 | Design | 4 |
| 9 | Documentation | 4 |
| 10 | Ethics | 5 |
| 11 | Infrastructure | 3 |
| 12 | Model | 13 |
| 13 | People | 3 |
| 14 | Process | 2 |
| 15 | Requirements | 5 |
| 16 | Self-Admitted (SATD) | 6 |
| 17 | Test | 4 |
| 18 | Versioning | 6 |
| | Total | 100 |

**Table 6. Selected studies by type of debt**

| Title | Author(s) | Year | Technical Debt Type |
|---|---|---|---|
| Algorithm Debt: Challenges and Future Paths | EIO Simon, M Vidoni, FH Fard | 2023 | Algorithm |
| Is using deep learning frameworks free? Characterizing and Measuring Technical Debt in Deep Learning Applications | Liu, Jiakun and Huang, Qiao and Xia, Xin and Shang, Weiyi | 2020 | Algorithm |
| Toward understanding deep learning framework bugs | Chen, J., Liang, Y., Shen, Q., Jiang, J., & Li, S. | 2023 | Algorithm |
| Understanding software-2.0: A study of machine learning library usage and evolution | Dilhara, M., Ketkar, A., & Dig, D. | 2021 | Algorithm |
| Adapting Software Architectures to Machine Learning Challenges | Serban, A. Visser, J. | 2022 | Architecture |
| An Empirical Study of Software Architecture for Machine Learning | Serban, A. Visser, J. | 2021 | Architecture |
| Architecting the Future of Software Engineering | Carleton, A., Shull, F., & Harper, E. | 2022 | Architecture |
| Architectural Decisions in AI-Based Systems: An Ontological View | Gómez, C. | 2022 | Architecture |
| Architecture Decisions in AI-based Systems Development: An Empirical Study | Zhang, B., Liu, T., Liang, P., Wang, C., Shahin, M., & Yu, J. | 2023 | Architecture |
| Code and Architectural Debt in Artificial Intelligence Systems | G Recupito, F Pecorelli, G Catolino et al. | 2024 | Architecture |
| Engineering AI Systems: A Research Agenda | Bosch, J., Olsson, H. H., & Crnkovic, I. | 2021 | Architecture |
| Software Architecture for ML-based Systems: What Exists and What Lies Ahead | Muccini, H., & Vaidhyanathan, K. | 2021 | Architecture |
| Software Engineering for AI-Based Systems: A Survey | Martínez-Fernández, S., Bogner, J., Franch, X., Oriol, M., Siebert, J., Trendowicz, A., ... & Wagner, S. | 2022 | Architecture |
| Studying Software Engineering Patterns for Designing ML Systems | Washizaki, H., Uchida, H., Khomh, F., & Guéhéneuc, Y. G. | 2019 | Architecture |
| Comprehending the Use of Intelligent Techniques to Support Technical Debt Management | Albuquerque, D., Guimaraes, E., Tonin, G., Perkusich, M., Almeida, H., & Perkusich, A. | 2022 | Build |
| Is using deep learning frameworks free? Characterizing and Measuring Technical Debt in Deep Learning Applications | Liu, Jiakun and Huang, Qiao and Xia, Xin and Shang, Weiyi | 2020 | Build |
| An Empirical Study of Refactorings and Technical Debt in Machine Learning Systems | Tang, Y., Khatchadourian, R., Bagherzadeh, M., Singh, R., Stewart, A., & Raja, A. | 2021 | Code |
| Better Code, Better Sharing: On the Need of Analyzing Jupyter Notebooks | Wang, J., Li, L., & Zeller, A. | 2020 | Code |
| Characterizing TD and Antipatterns in AI-Based Systems: A Systematic Mapping Study | Bogner, J., Verdecchia, R., & Gerostathopoulos, I. | 2021 | Code |
| Code and Architectural Debt in Artificial Intelligence Systems | G Recupito, F Pecorelli, G Catolino et al. | 2024 | Code |
| Code and Architectural Debt in Artificial Intelligence Systems | G Recupito, F Pecorelli, G Catolino et al. | 2024 | Code |
| Code Smells in Machine Learning Systems | Gesi, J., Liu, S., Li, J., Ahmed, I., Nagappan, N., Lo, D., ... & Bao, L. | 2022 | Code |
| How does machine learning change software development practices? | Wan, Z., Xia, X., Lo, D., & Murphy, G. C. | 2019 | Code |
| Software Quality for AI: Where We Are Now? | Lenarduzzi, V., Lomio, F., Moreschini, S., Taibi, D., & Tamburri, D. A. | 2021 | Code |
| Studying the Machine Learning Lifecycle and Improving Code Quality of Machine Learning Applications | Haakman, M. P. A. | 2020 | Code |
| The prevalence of code smells in machine learning projects | Van Oort, B., Cruz, L., Aniche, M., & Van Deursen, A. | 2021 | Code |
| An Empirical Study of Refactorings and Technical Debt in Machine Learning Systems | Tang, Y., Khatchadourian, R., Bagherzadeh, M., Singh, R., Stewart, A., & Raja, A. | 2021 | Configuration |
| Hidden Technical Debt in Machine Learning Systems | Sculley, D., Holt, G., Golovin, D., Davydov, E., Phillips, T., Ebner, D., ... & Dennison, D. | 2015 | Configuration |
| Software engineering challenges for machine learning applications: A literature review | Kumeno, F. | 2019 | Configuration |
| Software Engineering Challenges of Deep Learning | Arpteg, A., Brinne, B., Crnkovic-Friis, L., & Bosch, J. | 2018 | Configuration |
| Characterizing TD and Antipatterns in AI-Based Systems: A Systematic Mapping Study | Bogner, J., Verdecchia, R., & Gerostathopoulos, I. | 2021 | Data |
| Data Lifecycle Challenges in Production Machine Learning: A Survey | Polyzotis, N., Roy, S., Whang, S. E., & Zinkevich, M. | 2018 | Data |
| Data Smells: Categories, Causes and Consequences, and Detection of Suspicious Data in AI-based Systems | Foidl, H., Felderer, M., & Ramler, R. | 2022 | Data |
| Data Validation For Machine Learning | Polyzotis, N., Zinkevich, M., Roy, S., Breck, E., & Whang, S. | 2019 | Data |
| Software Quality for AI: Where We Are Now? | Lenarduzzi, V., Lomio, F., Moreschini, S., Taibi, D., & Tamburri, D. A. | 2021 | Data |
| Studying Software Engineering Patterns for Designing ML Systems | Washizaki, H., Uchida, H., Khomh, F., & Guéhéneuc, Y. G. | 2019 | Data |
| Technical Debt in Data-Intensive Software Systems | Foidl, H., Felderer, M., & Biffl, S. | 2019 | Data |
| The ML Test Score: A Rubric for ML Production Readiness and Technical Debt Reduction | Breck, E., Cai, S., Nielsen, E., Salib, M., & Sculley, D. | 2017 | Data |
| Towards Accountability for Machine Learning Datasets: Practices from Software Engineering and Infrastructure | Hutchinson, B., Smart, A., Hanna, A., Denton, E., Greer, C., Kjartansson, O., ... & Mitchell, M. | 2021 | Data |
| 23 Shades of Self-Admitted Technical Debt: an Empirical Study on Machine Learning Software | OBrien, D., Biswas, S., Imtiaz, S., Abdalkareem, R., Shihab, E., & Rajan, H. | 2022 | Defect |
| Characterizing TD and Antipatterns in AI-Based Systems: A Systematic Mapping Study | Bogner, J., Verdecchia, R., & Gerostathopoulos, I. | 2021 | Defect |
| Is using deep learning frameworks free? Characterizing and Measuring Technical Debt in Deep Learning Applications | Liu, Jiakun and Huang, Qiao and Xia, Xin and Shang, Weiyi | 2020 | Defect |
| Technical Debt Payment and Prevention Through the Lenses of Software Architects | Pérez, B., Castellanos, C., Correal, D., Rios, N., Freire, S., Spínola, R., ... & Izurieta, C. | 2021 | Defect |
| The ML Test Score: A Rubric for ML Production Readiness and Technical Debt | Breck, E., Cai, S., Nielsen, E., Salib, M., & Sculley, D. | 2017 | Defect |
| Reduction | | | |
| Common problems with Creating Machine Learning Pipelines from Existing Code | O'Leary, K., & Uchida, M. | 2020 | Design |
| Design Patterns for AI-based Systems: A Multivocal Literature Review and Pattern Repository | Heiland, L., Hauser, M., & Bogner, J. | 2023 | Design |
| Patterns and Anti-Patterns, Principles and Pitfalls: Accountability and Transparency in AI | Matthews, J. | 2020 | Design |
| Software-Engineering Design Patterns for Machine Learning Applications | Washizaki, H., Khomh, F., Guéhéneuc, Y. G., Takeuchi, H., Natori, N., Doi, T., & Okuda, S. | 2022 | Design |
| Characterizing TD and Antipatterns in AI-Based Systems: A Systematic Mapping Study | Bogner, J., Verdecchia, R., & Gerostathopoulos, I. | 2021 | Documentation |
| Collaboration Challenges in Building ML-Enabled Systems: Communication, Documentation, Engineering, and Process | Nahar, N., Zhou, S., Lewis, G., & Kästner, C. | 2022 | Documentation |
| Maintainability Challenges in ML: A Systematic Literature Review | Shivashankar, K., & Martini, A. | 2022 | Documentation |
| Understanding Implementation Challenges in Machine Learning Documentation | Chang, J., & Custis, C. | 2022 | Documentation |
| "This is Just a Prototype": How Ethics Are Ignored in Software Startup-Like Environments | Vakkuri, V., Kemell, K. K., Jantunen, M., & Abrahamsson, P. | 2020 | Ethics |
| Managing bias in AI | Roselli, D., Matthews, J., & Talagala, N. | 2019 | Ethics |
| Patterns and Anti-Patterns, Principles and Pitfalls: Accountability and Transparency | Matthews, J. | 2020 | Ethics |
| Principles alone cannot guarantee ethical AI | Mittelstadt, B. | 2019 | Ethics |
| Who pays for ethical debt in AI? | Petrozzino, C. | 2021 | Ethics |
| Characterizing TD and Antipatterns in AI-Based Systems: A Systematic Mapping Study | Bogner, J., Verdecchia, R., & Gerostathopoulos, I. | 2021 | Infrastructure |
| How does machine learning change software development practices? | Wan, Z., Xia, X., Lo, D., & Murphy, G. C. | 2019 | Infrastructure |
| Practices and Infrastructures for Machine Learning Systems: An Interview Study in Finnish Organizations | Muiruri, D., Lwakatare, L. E., Nurminen, J. K., & Mikkonen, T. | 2022 | Infrastructure |
| A Taxonomy of Software Engineering Challenges for Machine Learning Systems_ An Empirical Investigation | Lwakatare, L. E., Raj, A., Bosch, J., Olsson, H. H., & Crnkovic, I. | 2019 | Model |
| Clones in Deep Learning Code: What, where, and why? | Jebnoun, H., Rahman, M. S., Khomh, F., & Muse, B. A. | 2022 | Model |
| Empirical Analysis of Hidden Technical Debt Patterns in Machine Learning Software | Alahdab, M., & Çalıklı, G. | 2019 | Model |
| Exploring the Impact of Code Clones on Deep Learning Software | Mo, R., Zhang, Y., Wang, Y., Zhang, S., Xiong, P., Li, Z., & Zhao, Y. | 2023 | Model |
| Hidden Technical Debt in Machine Learning Systems | Sculley, D., Holt, G., Golovin, D., Davydov, E., Phillips, T., Ebner, D., ... & Dennison, D. | 2015 | Model |
| On Challenges in Machine Learning Model Management | Schelter, S., Biessmann, F., Januschowski, T., Salinas, D., Seufert, S., & Szarvas, G. | 2015 | Model |
| Quality issues in Machine Learning Software Systems | Côté, P. O., Nikanjam, A., Bouchoucha, R., Basta, I., Abidi, M., & Khomh, F. | 2023 | Model |
| Synergy Between Machine/Deep Learning and Software Engineering: How Far Are We? | Wang, S., Huang, L., Ge, J., Zhang, T., Feng, H., Li, M., ... & Ng, V. | 2020 | Model |
| The ML Test Score: A Rubric for ML Production Readiness and Technical Debt Reduction | Breck, E., Cai, S., Nielsen, E., Salib, M., & Sculley, D. | 2017 | Model |
| Towards CRISP-ML(Q): A ML Process Model with Quality Assurance Methodology | Studer, S., Bui, T. B., Drescher, C., Hanuschkin, A., Winkler, L., Peters, S., & Müller, K. R. | 2021 | Model |
| Towards Guidelines for Assessing Qualities of Machine Learning Systems | Siebert, J., Joeckel, L., Heidrich, J., Nakamichi, K., Ohashi, K., Namba, I., ... & Aoyama, M. | 2020 | Model |
| Understanding Development Process of Machine Learning Systems: Challenges and Solutions | de Souza Nascimento, E., Ahmed, I., Oliveira, E., Palheta, M. P., Steinmacher, I., & Conte, T. | 2019 | Model |
| What Is Really Different in Engineering AI-Enabled Systems? | Ozkaya, I. | 2020 | Model |
| Collaboration Challenges in Building ML-Enabled Systems: Communication, Documentation, Engineering, and Process | Nahar, N., Zhou, S., Lewis, G., & Kästner, C. | 2022 | People |
| How Do Engineers Perceive Difficulties in Engineering of Machine-Learning Systems? - Questionnaire Survey | Ishikawa, F., & Yoshioka, N. | 2019 | People |
| What is Social Debt in Software Engineering? | Tamburri, D. A., Kruchten, P., Lago, P., & van Vliet, H. | 2013 | People |
| Exploring the Impact of Code Clones on Deep Learning Software | Mo, R., Zhang, Y., Wang, Y., Zhang, S., Xiong, P., Li, Z., & Zhao, Y. | 2023 | Process |
| Studying Software Engineering Patterns for Designing ML Systems | Washizaki, H., Uchida, H., Khomh, F., & Guéhéneuc, Y. G. | 2019 | Process |
| A Taxonomy of Software Engineering Challenges for Machine Learning Systems_ An Empirical Investigation | Lwakatare, L. E., Raj, A., Bosch, J., Olsson, H. H., & Crnkovic, I. | 2019 | Requirement |
| It Takes Three to Tango: Requirement, Outcome/data, and AI Driven Development | Bosch, J., Olsson, H. H., & Crnkovic, I. | 2018 | Requirement |
| Requirements Engineering Challenges in Building AI-based Complex Systems | Belani, H., Vukovic, M., & Car, Ž. | 2019 | Requirement |
| Requirements Engineering for Artificial Intelligence Systems: A Systematic Mapping Study | Ahmad, K., Abdelrazek, M., Arora, C., Bano, M., & Grundy, J. | 2023 | Requirement |
| Requirements Engineering for Machine Learning: Perspectives from Data Scientists | Vogelsang, A., & Borg, M. | 2019 | Requirement |
| 23 Shades of Self-Admitted Technical Debt: an Empirical Study on Machine Learning Software | OBrien, D., Biswas, S., Imtiaz, S., Abdalkareem, R., Shihab, E., & Rajan, H. | 2022 | Self-Admitted (SATD) |
| A Large-Scale Empirical Study on Self-Admitted Technical Debt | Bavota, G., & Russo, B. | 2016 | Self-Admitted (SATD) |
| An Empirical Study of Self-Admitted Technical Debt in Machine Learning Software | Bhatia, A., Khomh, F., Adams, B., & Hassan, A. E. | 2023 | Self-Admitted (SATD) |
| Automating Change-level Self-Admitted Technical Debt Determination | Yan, M., Xia, X., Shihab, E., Lo, D., Yin, J., & Yang, X. | 2018 | Self-Admitted (SATD) |
| Is using deep learning frameworks free? Characterizing and Measuring Technical Debt in Deep Learning Applications | Liu, Jiakun and Huang, Qiao and Xia, Xin and Shang, Weiyi | 2020 | Self-Admitted (SATD) |
| Towards Automatically Addressing Self-Admitted Technical Debt: How Far Are We? | Mastropaolo, A., Di Penta, M., & Bavota, G. | 2023 | Self-Admitted (SATD) |
| A Systematic Mapping Study on Testing of Machine Learning Programs | Sherin, S., & Iqbal, M. Z. | 2019 | Test |
| Machine Learning Testing: Survey, Landscapes and Horizons | Zhang, J. M., Harman, M., Ma, L., & Liu, Y. | 2020 | Test |
| On Testing Machine Learning Programs | Braiek, H. B., & Khomh, F. | 2020 | Test |
| Testing Machine Learning based Systems: A Systematic Mapping | Riccio, V., Jahangirova, G., Stocco, A., Humbatova, N., Weiss, M., & Tonella, P. | 2020 | Test |
| "We Have No Idea How Models will Behave in Production until Production": How Engineers Operationalize Machine Learning | Shankar, S., Garcia, R., Hellerstein, J. M., & Parameswaran, A. | 2024 | Versioning |
| On Challenges in Machine Learning Model | Schelter, S., Biessmann, F., Januschowski, T., | 2015 | Versioning |
| Management | Salinas, D., Seufert, S., & Szarvas, G. | | |
| On the Challenges of Migrating to Machine Learning Life Cycle Management Platforms | Njomou, A. T., Fokaefs, M., Silatchom Kamga, D. F., & Adams, B. | 2022 | Versioning |
| Software Engineering Challenges of Deep Learning | Arpteg, A., Brinne, B., Crnkovic-Friis, L., & Bosch, J. | 2018 | Versioning |
| The Story in the Notebook_ Exploratory Data Science using a Literate Programming Tool | Kery, M. B., Radensky, M., Arya, M., John, B. E., & Myers, B. A. | 2018 | Versioning |
| Versioning for End-to-End Machine Learning Pipelines | Van Der Weide, T., Papadopoulos, D., Smirnov, O., Zielinski, M., & Van Kasteren, T. | 2017 | Versioning |

### A.9.1 Key Types of Technical Debt Identified

#### A.9.1.1 Algorithm

**Definition and Scope**

Algorithm Debt (AD) is a specific type of technical debt (TD) that arises from sub-optimal implementations of algorithm logic, particularly in performance-critical and algorithm-intensive projects such as AI frameworks and scientific software. Similar to other forms of TD, AD results in increased maintenance burdens, potential defects, and reduced system performance over time.

**Key Characteristics**

*Sub-optimal Implementations.*

Algorithm debt often involves inefficient or non-optimized algorithms that negatively impact the performance and scalability of AI systems. Examples include poorly implemented pooling methods in neural networks or unoptimized cost functions.

*Impact on AI and Scientific Software*

Given the critical nature of AI-driven applications in fields such as finance, healthcare, and transportation, the presence of AD can lead to significant real-world consequences. Notable incidents include financial losses due to algorithmic trading errors and safety failures in autonomous vehicle systems.

*Challenges in Detection and Management*

AD is often identified through self-admitted technical debt (SATD) comments left by developers, highlighting the known sub-optimal aspects of the code that need improvement. The evolving nature of AI algorithms and the rapid pace of development exacerbate the difficulty in addressing AD, as new techniques and optimizations are continuously emerging.

**Distribution in Deep Learning Frameworks**

Studies indicate that AD constitutes a significant portion of the overall technical debt in deep learning frameworks, with its prevalence varying across different projects:

Design debt is the most common, followed by requirement debt and algorithm debt, which can range from 5.62% to 20.67% of the total TD instances.

The distribution of AD highlights the ongoing struggle to implement cutting-edge algorithms efficiently and the need for continuous optimization.

**Economic and Safety Impacts**

Economic Impact: The case of Knight Capital, which suffered massive financial losses due to rushed and faulty algorithm implementations, underscores the economic risks associated with AD.

Safety Impact: AD in critical systems, such as autonomous vehicles, can lead to severe safety incidents, including fatalities, as demonstrated by the Uber self-driving car accident in Arizona.

**Management Strategies**

*Collaboration and Expertise*

Encouraging collaboration among developers and seeking expert advice can help mitigate the risks associated with AD.

Regular code reviews and refactoring sessions can identify and address sub-optimal algorithms early in the development process.

*Testing and Documentation*

Comprehensive testing frameworks and thorough documentation are essential to ensure that algorithmic implementations are robust and well-understood by all team members.

Using multiple frameworks to test AI algorithms can help uncover AD and improve overall algorithm performance and reliability.

*Awareness and Training*

Increasing awareness of AD among developers and providing training on best practices for algorithm implementation can reduce the likelihood of incurring AD.

**Future Research Directions**
Further research is needed to develop standardized methods for detecting and measuring AD.
Investigating the generality of AD across different types of AI and scientific software can provide insights into common patterns and solutions.
Algorithm Debt represents a significant challenge in the development and maintenance of AI-driven systems, necessitating focused efforts on optimization, collaboration, and continuous improvement to ensure high-quality and reliable software.

#### A.9.1.2 Architecture

**Definition and Scope**
Architectural debt refers to suboptimal or compromised architectural choices made during software development to achieve short-term benefits at the cost of future challenges and increased maintenance. This concept parallels financial debt, where "interest" must be paid over time in the form of additional work required to address the consequences of these early decisions. In the context of AI-enabled systems, architectural debt can have profound implications due to the complexity and unique requirements of these systems.

**Key Aspects of Architectural Debt**
*Prevalence and Impact*
Architectural debt is prevalent in AI-based systems due to the rapid evolution and complexity of integrating AI components into existing architectures. Issues such as jumbled model architectures and pipeline jungles are common, where the components are difficult to understand, maintain, and evolve.
*Causes*
Short-term Solutions: Decisions made to quickly address immediate needs without considering long-term impacts.
Integration Challenges: Integrating AI components often requires significant changes to the existing architecture, leading to increased debt.
Data Quality and Management: Poor data quality and management practices contribute significantly to architectural debt, as AI systems heavily rely on data integrity and consistency.
*Types of Architectural Debt*
Pipeline Jungle: Evolving data pipelines without adequate management can lead to a complex, unmanageable system.
Jumbled Model Architecture: AI models cobbled together in an ad hoc manner, leading to maintenance difficulties.
Technical Debt in Model Code: Issues specific to the model code, such as overfitting, lack of modularization, and difficulties in debugging.
*Consequences*
Increased maintenance costs and effort.
Difficulty in scaling and evolving the system.
Compromised system performance and reliability.
*Management Strategies*
Regular Refactoring: Continual refactoring of the architecture to address emerging issues and prevent debt accumulation.
Automated Testing and Monitoring: Implementing comprehensive testing and monitoring frameworks to catch and address issues early.
Documentation and Best Practices: Maintaining thorough documentation and adhering to best practices to ensure that architectural decisions are made with long-term considerations in mind.
Empirical Studies and Surveys: Conducting studies to understand the prevalence, severity, and mitigation strategies from the perspective of developers and practitioners.
*Future Directions*
Tool Development: Creating tools to assist in the identification, evaluation, and management of architectural debt in AI systems.
Research and Methodologies: Further research into methodologies that can help manage and reduce architectural debt, especially in the context of AI and machine learning.
In conclusion, architectural debt in AI-enabled systems poses significant challenges but can be managed through proactive strategies, continuous improvement, and the adoption of best practices. Addressing architectural debt is crucial for maintaining the long-term health and sustainability of AI systems.

#### A.9.1.3 Build

**Definition and Scope**
Build debt is a type of technical debt that arises during the software development process due to suboptimal decisions made for the sake of expediency, such as shortening the completion time, reducing costs, or meeting market demands. These decisions often result in a trade-off where immediate benefits are gained at the expense of the long-term health of the software system. Build debt accumulates as a result of various forms of technical debt, including design debt, defect debt, documentation debt, requirement debt, and test debt, among others.

**Management Strategies**
Effective management of build debt involves a set of practices aimed at identifying, measuring, monitoring, and resolving technical debt. These practices include:
Regular Refactoring: Continuously improving the codebase by addressing design debt and refactoring suboptimal code.
Enhanced Testing: Increasing test coverage and ensuring that tests are updated and relevant to catch defects early.
Comprehensive Documentation: Maintaining up-to-date and thorough documentation to facilitate understanding and collaboration among developers.
Prioritizing Debt Repayment: Strategically planning and prioritizing the repayment of technical debt based on its impact on the project and the resources available.

**Role of Intelligent Techniques**
Intelligent techniques, such as machine learning, natural language processing, and expert systems, can support technical debt management by:
Identifying Technical Debt: Using data analysis and pattern recognition to detect instances of technical debt in the codebase.
Supporting Decision-Making: Providing insights and recommendations on the prioritization and resolution of technical debt.
Automating Tasks: Automating repetitive tasks such as code refactoring and testing to reduce the manual effort required to manage technical debt.
By leveraging these techniques, organizations can enhance their ability to manage build debt effectively, ensuring the long-term health and sustainability of their software systems.

### A.9.1.4 Code

**Definition and Scope**
Code debt is a form of technical debt that specifically refers to suboptimal code that requires additional effort to maintain, understand, and evolve over time. It includes issues like code smells, duplicate code, and poorly organized code structures. In AI and machine learning (ML) systems, code debt can have significant impacts due to the complexity and rapid evolution of these systems.

**Common Issues in ML Systems**
*Code Duplication*
Prevalence: Code duplication is one of the most significant issues in ML systems. It mainly occurs in ML configuration and model code, where similar algorithms and configurations are frequently replicated.
Impact: This leads to increased maintenance efforts, higher risk of bugs, and challenges in implementing updates consistently.
*Configuration Debt*
Description: Configuration debt involves poorly managed or scattered configuration settings for ML models and algorithms.
Impact: It complicates the tuning and updating of models, leading to potential errors and inefficiencies .
*Dead Experimental Code Paths*
Description: These are remnants of prototype code that were used for testing new features or algorithms but were not removed after the features were finalized.
Impact: Such paths clutter the codebase, making it harder to maintain and increasing the likelihood of introducing errors .
*Plain-Old-Data Types (PODT) Debt*
Description: Use of simple data types instead of richer, more descriptive types that better encapsulate the information and functionalities required.
Impact: This leads to poor readability, harder maintenance, and increased potential for bugs due to lack of context and structure in the data representation.
*Model Code Debt*
Description: Issues specific to the code implementing ML models, including poor modularization and lack of abstraction.
Impact: This affects the ability to update and extend models, making the system less adaptable to new requirements or improvements.

**Strategies for Managing Code Debt**
*Refactoring*
Approaches: Common refactorings include eliminating duplicate code through inheritance, improving modularization, and enhancing code clarity and organization.
Benefits: These practices help in reducing the overall complexity, improving maintainability, and enhancing the clarity of the codebase.
*Static Code Analysis*

Tools: Use of tools like Pylint to identify code smells and other potential issues early in the development process.
Benefits: Early detection and resolution of issues can significantly reduce the long-term maintenance burden and improve code quality.

*Adherence to Best Practices*
Recommendations: Adopting best practices such as using descriptive variable names, favoring polymorphism over flags for algorithm variations, and properly organizing configuration parameters.
Impact: These practices lead to a more organized, readable, and maintainable codebase, reducing the likelihood of accumulating code debt.

**Conclusion**
Addressing code debt in ML systems is critical for maintaining high-quality, efficient, and adaptable software. By recognizing common issues and implementing strategic refactorings and best practices, developers can mitigate the negative impacts of code debt, ensuring the long-term sustainability and performance of their ML applications.

### A.9.1.5 Configuration

**Definition and Scope**
Configuration debt is a significant type of technical debt in machine learning (ML) systems. It arises from the complexities involved in managing the myriad of configurable options within ML models and systems, such as feature selection, data handling methods, algorithm-specific settings, and preprocessing and postprocessing steps.

**Characteristics and Issues**
*Complexity and Error Potential*
Large ML systems have numerous configurable parameters, each with the potential for mistakes. Misconfigurations can lead to substantial issues, such as degraded model performance, wasted computational resources, and increased maintenance costs .
*Lack of Verification and Quality Control*
Configuration changes are often treated as an afterthought, and their verification or testing may not be given the same level of importance as code changes. This oversight can result in configurations that are hard to modify correctly and reason about .
*Impact on Model Performance and Size*
Configuration settings can significantly impact the performance of ML models. In mature systems, the number of lines of configuration can exceed the number of lines of traditional code, each line introducing potential errors .
*Examples of Configuration Issues*
Incorrectly logged features, unavailable data for certain periods, changes in data logging formats, and the need for different features in production versus training environments are common examples. These issues complicate configuration management and can lead to inefficiencies and errors in ML systems .

**Principles for Good Configuration Systems**
*Ease of Specification and Modification*
Configurations should be easy to specify and modify, ideally allowing small changes from previous configurations.
*Error Prevention*
The system should minimize the potential for manual errors, omissions, or oversights.
*Visual Comparison*
It should be easy to visually compare different configurations to identify changes and discrepancies.
*Automated Assertions and Verification*
Systems should support automated assertions and verifications of basic facts about configurations, such as the number of features used and data dependencies.
*Detection of Redundancy:*
The system should detect unused or redundant settings to streamline configurations.
*Code Review and Repository Integration*
Configuration changes should undergo thorough code reviews and be checked into a repository to maintain version control and traceability.

**Addressing Configuration Debt**
Refactoring efforts often focus on reducing configuration debt through various strategies:
*Reorganization and Inheritance*
- Introducing inheritance and reorganizing class hierarchies can centralize configuration management, making it easier to handle multiple learning algorithm variations and reducing redundant configuration code.
*Elimination of Duplication*
- Efforts to eliminate duplicate code in configuration settings and model implementations can simplify maintenance and reduce the likelihood of errors.

**Conclusion**

Configuration debt represents a major challenge in the maintenance and evolution of ML systems due to the intricate and error-prone nature of managing extensive configurable options. Effective strategies to mitigate this debt include enhancing the ease of configuration modifications, preventing manual errors, automating verification processes, and adopting best practices in refactoring to streamline and centralize configuration management.

### A.9.1.6 Data

Data Debt is a specific type of technical debt unique to AI-based systems, referring to deficiencies in the collection, management, and usage of data used in these systems. It is identified as a significant concern due to the heavy reliance of AI on data. The main issues associated with data debt include:

*Data Quality Issues*

Poor quality data can lead to reduced classification effectiveness and inaccurate model predictions. This encompasses errors in data collection, noisy or incomplete datasets, and lack of proper data cleaning mechanisms.

*Unmanaged Data Dependencies and Anomalies*

AI systems often have complex data pipelines with multiple dependencies. Poorly managed data dependencies can result in broken pipelines, leading to failures in data processing and model training.

*Data Relevance*

Using irrelevant or outdated data can degrade model performance. Ensuring that data remains relevant and up-to-date is crucial for maintaining the accuracy and reliability of AI systems.

*Latent Data Debt*

Some issues related to data debt may not manifest immediately but pose long-term risks to the system’s evolution. This can include legacy data practices that are hard to change and accumulate over time, making future maintenance and upgrades challenging.

*Compatibility Issues*

Inconsistencies between different datasets and data formats can lead to integration problems, affecting the seamless operation of AI models and their deployment.

The impact of data debt is profound, as it can lead to reduced classification effectiveness, data loss due to premature aggregation, and other compatibility issues. It emphasizes the need for robust data management practices, continuous monitoring, and proactive measures to address data-related challenges in AI-based systems.

### A.9.1.7 Defect

Defect debt refers to code that does not function as intended but whose repair is postponed due to factors such as repair complexity or time pressure. This type of debt is particularly prevalent in deep learning frameworks, where the non-deterministic nature of probability and statistics-based algorithms complicates defect reproducibility. For instance, bugs may be rare or complex, making immediate fixes challenging and often leading to defect debt.

**Key characteristics of defect debt include**

*Non-deterministic Issues*

The stochastic behavior of machine learning algorithms can lead to bugs that are hard to replicate and fix.

*Inter-team Dependencies*

Delays in addressing defects often stem from the need for collaboration among different development teams. This coordination can be time-consuming and contribute to the accumulation of defect debt.

*Impact on Quality and Development*

Accumulated defect debt can degrade the quality of frameworks, leading to unstable data dependencies. Application developers might unknowingly base their models on flawed frameworks, resulting in unexpected outcomes and further complicating future development.

**Examples from deep learning frameworks illustrate the challenges**

TensorFlow developers have acknowledged potential infinite loops due to small maximum tensor sizes, which are postponed for repair due to the complexity involved.

MXNet developers delay fixes for issues that require confirmation from other teams, such as core dumps caused by specific operations.

**Conclusion**

In summary, defect debt in machine learning systems can significantly hinder development efficiency and system reliability, necessitating careful management to prevent long-term detrimental effects on software quality and maintainability.

### A.9.1.8 Design

**Definition and Concept of Design Debt**

Design debt, often interlinked with technical debt, refers to suboptimal design decisions that may expedite development in the short term but result in increased complexity and maintenance costs in the long term. This debt is particularly critical in AI-based systems where the complexity and evolving nature of machine learning models amplify its impact.

**Sources and Manifestations of Design Debt**

*Monolithic Program Structures*: AI systems, especially those developed without modular design practices, often result in monolithic programs. These systems lack clear boundaries and separation of concerns, leading to difficulties in maintenance and scalability. The absence of modularization hinders the reusability of components and the integration of new functionalities, contributing significantly to design debt.

*Ad-Hoc Problem Solving*: A substantial portion of machine learning practitioners address design challenges in an ad-hoc manner without adhering to established design patterns. This approach may solve immediate issues but accumulates design debt due to the lack of a systematic and reusable framework.

*Complexity and Quality Concerns*: The inherent complexity of machine learning systems, combined with concerns about software quality, often leads to design debt. Inadequate documentation, lack of standardization in model implementation, and insufficient validation processes contribute to this debt.

**Addressing Design Debt through Design Patterns**

Design patterns provide reusable solutions to common design problems and can significantly mitigate design debt. By promoting best practices and standardization, design patterns help in managing the complexity and improving the maintainability of AI-based systems.

*Pattern Adoption*: Despite the benefits, the adoption of design patterns is not widespread. Studies have shown that only a minority of developers consistently use design patterns, while others either rely on external documented patterns or resolve problems in an ad-hoc way.

*Categories of Patterns*

Topology Patterns: Address the overall system architecture, ensuring that different workloads and components are properly isolated and managed.

Programming Patterns: Focus on the design of specific components within the system, promoting modularity and encapsulation.

Model Operation Patterns: Concerned with the management and operation of machine learning models, ensuring robustness, explainability, and accuracy.

*Examples of Patterns*

Encapsulating ML Models within Rule-based Safeguards. This pattern introduces deterministic, rule-based mechanisms to handle the predictions of ML models, thereby mitigating risks associated with incorrect predictions and enhancing system reliability.

Workflow Pipeline. This pattern involves creating containerized services for each pipeline step, which improves the portability, scalability, and maintainability of the pipeline at the cost of increased complexity.

**Conclusion**

Managing design debt in AI-based systems requires a systematic approach to design, leveraging established design patterns to promote modularity, scalability, and maintainability. While the adoption of these patterns is still not universal, their benefits in reducing design debt and improving software quality are well-documented. Raising awareness and making these patterns more accessible to practitioners are crucial steps towards mitigating design debt in AI systems.

### A.9.1.9 Documentation

Documentation debt, a subset of technical debt, refers to the deficiencies and shortcuts in the documentation practices within AI-based systems that can lead to increased maintenance costs, reduced system comprehensibility, and hindered collaboration and transparency. It shares commonalities with other types of software systems but also exhibits unique challenges and characteristics in the context of AI and machine learning (ML).

**Key Characteristics and Challenges**

*Incomplete or Missing Documentation*

Documentation debt often arises from incomplete or missing documentation, which makes it difficult for current and future developers to understand the system's design, decisions, and changes.

*Impact on Quality Attributes*

Documentation debt can significantly impact the maintainability and reproducibility of AI systems. Poor documentation practices can lead to difficulties in modifying and testing AI models and systems, ultimately affecting their reliability and performance.

*Variations in Documentation Debt*
In AI-based systems, documentation debt can extend to features and assumptions on the used data. This includes undocumented data dependencies, annotations, and transformations that are crucial for understanding model behavior and performance.
*Challenges in Implementation*
Practitioners often view documentation as a tedious and time-consuming task. The process of gathering scattered information across various sources within an organization adds to this perception. Additionally, the high turnover of project owners can lead to the loss of critical knowledge about models and data, exacerbating documentation debt.
*Need for Standardization*
There is a recognized need for standardized documentation frameworks and practices to mitigate documentation debt. These frameworks should provide clear guidelines on what information needs to be documented and how it should be structured to ensure consistency and completeness.

**Recommendations**
*Collaborative Documentation Efforts*
Engaging non-technical stakeholders such as project managers, business analysts, and technical writers can help in reducing the burden on engineers and ensure that documentation is comprehensive and accessible to various audiences.
*Iterative Documentation Processes*
Implementing iterative feedback loops from documentation consumers can help ensure that the documentation remains relevant, accurate, and useful. This approach can also help in identifying missing information and improving the overall quality of documentation.
*Automated Documentation Tools*
The development and integration of automated documentation tools can help capture relevant information throughout the ML lifecycle, reducing the manual effort required and ensuring that documentation is up-to-date and complete.

**Conclusion**
Addressing documentation debt is crucial for maintaining the long-term sustainability and transparency of AI-based systems. By implementing structured documentation practices and leveraging collaborative and automated approaches, organizations can mitigate the risks associated with documentation debt and improve the overall quality and maintainability of their AI solutions.

### A.9.1.10 Ethics

Ethics Debt in AI refers to the accumulation of ethical issues and consequences that arise when AI systems are designed, developed, and deployed without adequately addressing ethical concerns. This concept is analogous to technical debt, where shortcuts and suboptimal solutions are initially chosen to save time or resources, with the expectation of addressing these issues later.

**Key Points**
*Definition and Nature*
Ethics Debt occurs when AI solutions are implemented without a thorough examination of their ethical implications, such as fairness, accountability, and transparency.
Unlike technical debt, which involves deliberate decisions to defer certain technical fixes, ethical debt often arises from the assumption that the AI solution is inherently ethical.
*Impact and Consequences*
Ethical debt can lead to significant harm, especially to vulnerable populations, by perpetuating biases and unfair practices.
For example, AI systems used in healthcare and hiring have been shown to exhibit biases that disadvantage minorities and women, leading to unequal treatment and opportunities.
*Detection and Challenges*
Ethical issues in AI can often only be detected after deployment, making them more challenging to address. Problems such as AI drift, where the model's performance degrades over time, can exacerbate these issues.
The opacity of AI systems, particularly those based on deep learning, makes it difficult for users and even developers to understand how decisions are made, further complicating the identification and resolution of ethical problems.
*Responsibility and Accountability*
The misalignment between those who incur ethical debt (developers and companies) and those who pay for it (users and affected individuals) is a significant issue. Ethical debt often burdens those who have the least ability to address or mitigate its impacts.
Existing professional and organizational frameworks often lack the necessary mechanisms to enforce ethical standards and hold developers accountable for ethical lapses.
*Mitigation and Best Practices*

Proactively addressing ethical concerns throughout the AI lifecycle – from ideation through to deployment and monitoring – is crucial to mitigating ethical debt.
Incorporating ethical analysis into the development process, involving diverse stakeholders, and creating robust oversight mechanisms are essential steps towards reducing ethical debt.

**Conclusion**
Ethical debt in AI represents a significant challenge that parallels the well-known concept of technical debt. It arises from insufficient attention to ethical considerations during the development and deployment of AI systems. Addressing this debt requires ongoing commitment, transparency, and the implementation of robust ethical frameworks to ensure that AI technologies serve all segments of society fairly and responsibly.

### A.9.1.11 Infrastructure

Infrastructure debt, within the context of AI-based systems, represents the technical debt associated with the infrastructure required to implement and operate these systems. This debt type is extended to include specific challenges related to AI pipelines and model management. Here is a detailed overview based on the extracted information:

**Definition**
Infrastructure debt refers to deficiencies in the infrastructure needed to build, deploy, and maintain AI systems. This includes issues with the underlying hardware, software, and network resources that support AI pipelines and model operations.

**Characteristics and Challenges**
*Complex Infrastructure*: AI systems often require sophisticated and interconnected pipelines for data processing, training, and deployment. These pipelines can become complex, resulting in "pipeline jungles" where components are precariously linked together via glue code.
*Resource Allocation*: Managing resources efficiently to train and test AI models is a significant challenge. Suboptimal resource allocation can lead to increased costs and inefficiencies.
*Monitoring and Debugging*: Effective monitoring and debugging of AI models are critical but often deficient. AI components necessitate additional observability requirements, such as monitoring data sources and model accuracy. The black-box nature of AI models complicates root cause analysis without specialized tools.
*Reproducibility*: Ensuring that AI models and their results are reproducible is a key aspect of infrastructure debt. This involves managing and versioning data and models adequately, which is often neglected, leading to reproducibility issues.

**Impact**
Infrastructure debt can severely affect the maintainability, performance, and reliability of AI systems. It often leads to increased operational costs and can make future changes or improvements to the system more difficult and costly.

**Solutions**
To address infrastructure debt, several strategies can be implemented:
*Manage Model Configuration*: Track, review, and test configuration changes in AI applications similarly to code changes. Externalize configuration options from the code to maintain them in human- and machine-readable files.
*Clear Component and Code APIs*: Reduce design and architectural debt by encapsulating AI functionality in well-defined software components with clear interfaces. This helps manage the glue code antipattern and makes the system more modular and maintainable.
*Monitor Deployed Models*: Continuously monitor AI models and their prediction performance post-deployment to detect and address issues like training/serving skew.

**Conclusion**
Infrastructure debt in AI-based systems represents a significant challenge due to the unique requirements of AI pipelines and model management. Addressing this debt involves implementing robust configuration management, clear APIs, and continuous monitoring to ensure system reliability and maintainability.

### A.9.1.12 Model

Model Debt is a subset of technical debt specific to machine learning (ML) systems, reflecting the additional complexity and maintenance challenges introduced by the unique properties and dependencies of ML models. It encompasses a variety of issues that arise from the interactions between data, model behavior, and system-level considerations. Here are the key points drawn from the provided documents:

*Concept and Importance*

ML systems often incur significant ongoing maintenance costs despite their quick initial development benefits. These costs arise from various ML-specific risk factors that extend beyond traditional software engineering problems.

*Categories of Model Debt*

Boundary Erosion: ML models can silently erode abstraction boundaries, making it difficult to maintain clear interfaces between system components. This erosion can result from the re-use or chaining of input signals, creating unintended dependencies between otherwise disjoint systems.

*Entanglement*: The "CACE principle" (Changing Anything Changes Everything) highlights how changes in one part of the model (e.g., input distributions, features) can affect the entire system, making isolation of improvements challenging.

*Feedback Loops*: Models that influence their own training data over time can create direct and hidden feedback loops, complicating predictions and system behavior.

*Undeclared Consumers*: Predictions made by ML models can be consumed by other systems without proper documentation or access controls, leading to hidden dependencies and feedback loops that increase maintenance difficulty.

*Data Dependencies*: ML models are highly sensitive to the data they consume. Unstable data dependencies and underutilized data dependencies can significantly impact model performance and maintainability.

*Configuration Debt*: The numerous configurable options in ML systems (e.g., features used, algorithm settings) can lead to complex and error-prone configurations that are hard to manage and maintain.

**System-Level Anti-Patterns**

*Glue Code and Pipeline Jungles*: ML systems often require extensive supporting code to interface with general-purpose packages, leading to complex and brittle pipeline jungles for data preparation and feature extraction.

*Dead Experimental Codepaths*: Experimental codepaths, if not managed properly, can accumulate over time, increasing system complexity and maintenance burden.

*Abstraction Debt*: A lack of strong abstractions in ML systems can blur the lines between components, making it harder to manage and improve systems over time.

**Strategies for Managing Model Debt**

*Early Detection*: Identifying and addressing model debt patterns during the prototyping phase is crucial to prevent them from becoming entrenched in the system.

*Holistic Design*: Considering data collection, feature extraction, and system integration holistically can reduce the complexity and cost of maintaining ML systems.

*Testing and Monitoring*: Implementing comprehensive testing and monitoring strategies can help detect and mitigate model debt, ensuring the system behaves as intended over time.

**Conclusion**

Understanding and managing model debt is essential for maintaining the long-term health and performance of ML systems. By recognizing the unique challenges posed by ML-specific issues and employing strategic design, testing, and monitoring practices, practitioners can mitigate the compounding costs of model debt.

### A.9.1.13 People

People Debt, as discussed in the context of software engineering, is synonymous with Social Debt. It refers to the unforeseen project costs associated with a suboptimal development community. This can arise from factors such as organizational barriers, global distance, and poor socio-technical decisions that affect both social and technical aspects of software development. Social Debt impacts the success of software projects significantly, similar to Technical Debt, but it is related to the people and their interactions rather than the codebase.

**Causes of People Debt**

*Socio-Technical Decisions*

Changes in the structure of the development community (e.g., outsourcing).

Adoption of new development processes (e.g., agile methods).

Leveraging global collaboration which requires balancing formal and informal communication.

*Organizational Dynamics*

Lack of trust within the community.

Immaturity or inability to tackle development problems.

Changes in team composition without proper handovers or documentation.

*Community Features and Defects*

Visible community features can have predictable impacts, whereas defects often result in increased Social Debt over time.

Scenarios from practice reveal that socio-technical decisions produce both visible and invisible effects on the community, some of which may contribute to Social Debt if not managed properly.

**Impacts of People Debt**
*Development Delays*
Social Debt can lead to inefficiencies and delays in software development due to strained social relationships and poor community structures.
*Increased Costs*
Like Technical Debt, Social Debt incurs additional costs that may not be immediately visible but increase over time as projects progress and issues become more complex to resolve.
*Quality Issues*
Poor community interactions can affect the overall quality of the software product, leading to bugs, poor performance, and ultimately, customer dissatisfaction.

**Management and Mitigation**
Formalizing Socio-Technical Decisions and providing mechanisms to detect and manage Social Debt by formalizing socio-technical decisions and measuring their impact on the development community.
*Awareness and Training*
Investing in training sessions, clear documentation, and guidelines to help teams understand the socio-technical aspects of their work and reduce the potential for Social Debt.
*Continuous Improvement*
Adopting continuous improvement practices and maintaining a dynamic approach to community management to ensure that socio-technical decisions do not lead to long-term debt.

**Conclusion**
In conclusion, People Debt is a crucial aspect of software engineering that parallels Technical Debt but focuses on the social dynamics within development teams. Understanding and managing this debt is essential for ensuring the success and sustainability of software projects.

### A.9.1.14 Process

Process debt refers to the accumulation of inefficiencies, waste, and redundancies within workflows over time. This concept is similar to technical debt but focuses on processes rather than code. Process debt can emerge due to outdated or suboptimal procedures that are no longer effective but remain in use because they are embedded in the organization's operations.

**Causes of Process Debt**
*Company Culture*: A resistance to change and reliance on outdated methods often lead to process debt. Organizations that stick to "how things have always been done" without regularly reviewing and updating their processes are more susceptible.
*Capacity and Time*: Limited resources and time constraints prevent organizations from addressing inefficient processes. When there is a lack of prioritization for process improvements, process debt continues to grow.
*Technological Advancements and Changes*: As tools and technologies evolve, processes must adapt. However, if new tools are integrated without revisiting and optimizing existing processes, inefficiencies can accumulate.

**Impact of Process Debt**
*Reduced Efficiency*: Inefficient processes slow down operations, leading to delays and increased costs.
*Employee Frustration*: Redundant and cumbersome processes can lead to dissatisfaction and lower productivity among employees.
*Hindered Growth*: Process debt can consume resources that could otherwise be used for innovation and growth.

**Mitigation Strategies**
*Regular Audits*: Regularly reviewing and auditing processes from start to finish can help identify bottlenecks and inefficiencies.
*Documentation*: Keeping detailed and up-to-date documentation of processes ensures that everyone understands the workflow and can identify areas for improvement.
*Prioritization and Capacity Management*: Allocating time and resources to process improvements and making it a priority can help mitigate process debt over time.

**Conclusion**
By understanding and addressing process debt, organizations can improve their efficiency and overall performance.

### A.9.1.15 Requirements

Requirements Debt refers to the accumulation of suboptimal decisions and deferred improvements related to requirements engineering, particularly in the context of machine learning (ML) systems. This concept is akin to technical debt, where the cost of rework increases due to shortcuts or compromises made during the requirements engineering process. In ML systems, requirements debt can arise from various factors including entanglement, undeclared consumers, and data dependencies.

*Entanglement*: ML models inherently create complex dependencies where changes in one part of the model can affect other parts unpredictably. This makes the isolation of improvements difficult and contributes to the CACE (Changing Anything Changes Everything) principle, increasing the requirements debt by complicating future modifications.
*Undeclared Consumers*: Predictions from ML models can be consumed by multiple systems, either at runtime or via logs. When these predictions are used as inputs for other systems without being clearly declared, it leads to hidden dependencies that are difficult to manage and track, thus adding to the requirements debt.
*Data Dependencies*: These dependencies can be unstable, underutilized, or difficult to analyze statically. For example, unstable dependencies change behavior over time, and underutilized dependencies include input features that add little value but remain part of the system. Correction cascades, where improving a model based on earlier solutions becomes increasingly costly, also contribute significantly to requirements debt.
*Data Requirements*: In ML systems, data requirements need to be specified meticulously to ensure quality and reliability. This includes identifying the necessary quantity and quality of data, handling imbalanced datasets, and ensuring the integrity and provenance of data sources. Failing to do so can lead to significant requirements debt as poor data quality or inadequate data can compromise the performance and fairness of the ML system.
*Non-functional Requirements (NFRs):* Traditional approaches to NFRs often fall short in ML contexts. For example, requirements related to fairness, transparency, and explainability are critical for ML systems but are frequently neglected or inadequately addressed. This oversight contributes to requirements debt, as addressing these NFRs retrospectively is often more complex and costly.
*Legal and Regulatory Requirements*: Compliance with legal and regulatory frameworks, such as GDPR, is essential in ML systems. Requirements engineers must ensure that personal data is used in ways that comply with legal standards. Failure to address these requirements from the outset can lead to significant debt, as retrofitting compliance measures can be highly challenging and costly.

**Conclusion**
In summary, requirements debt in ML systems is a critical issue that arises from the inherent complexity and interdependencies of these systems. Addressing this debt requires meticulous requirements engineering practices that consider the unique challenges of ML, including data quality, entanglement, and non-functional requirements. By proactively managing these aspects, requirements engineers can mitigate the accumulation of requirements debt and ensure the long-term maintainability and reliability of ML systems.

### A.9.1.16 Self-Admitted (SATD)

Self-Admitted Technical Debt (SATD) refers to instances where developers explicitly acknowledge the presence of technical debt in their code through comments. This concept helps in understanding and managing technical debt more effectively.

**Key Characteristics of SATD**
*Intentional Acknowledgment*
SATD is characterized by developers' intentional acknowledgment of suboptimal solutions or temporary workarounds in code comments.
*Types of SATD*
Code Debt: Suboptimal code that needs improvement.
Defect Debt: Known bugs or defects that are yet to be fixed.
Design Debt: Poor design choices that hinder future development.
Requirement Debt: Trade-offs or compromises made in meeting requirements.
Documentation Debt: Inadequate or missing documentation.
Test Debt: Insufficient testing or test coverage.
*Prevalence in Software Projects*
SATD is prevalent in many software systems, with an average of 51 instances per system. It tends to persist for a long time, often spanning over 1,000 commits before being addressed.
*Evolution and Removal*
SATD instances increase over time as new instances are introduced more frequently than they are removed. Around 57% of SATD instances are eventually fixed, but many remain unresolved for extended periods.
*Impact on Software Quality*

SATD can make code more difficult to maintain and evolve. It does not necessarily correlate with poor code quality but indicates areas needing improvement.

*Detection and Management*

SATD is identified by mining code comments for patterns indicating technical debt, such as "TODO," "FIXME," and "HACK." Automated tools and manual reviews help in identifying and managing SATD.

*Machine Learning Software*

In ML software, SATD is particularly common in components like data preprocessing and model building. Unique types of debt in ML include configuration debt and inadequate tests, reflecting the complexities of ML pipelines.

**Practical Implications**

*For Developers*

Acknowledge technical debt explicitly in code comments to facilitate future maintenance.

Prioritize fixing long-standing SATD to reduce its negative impact on software quality.

*Managers*

Implement regular reviews of code comments to identify and address SATD.

Provide resources and time for developers to address acknowledged technical debt.

*For Researchers*

Explore automated techniques to better detect and manage SATD.

Investigate the long-term impact of SATD on software projects and propose best practices for its management.

**Conclusion**

Understanding and managing SATD is crucial for maintaining high-quality software and ensuring the long-term sustainability of software projects.

## A.9.1.17 Test

Test debt, a concept derived from technical debt, refers to the accumulated cost of suboptimal testing practices in software development. In the context of machine learning (ML), test debt encompasses the inefficiencies and potential pitfalls associated with inadequate or improper testing of ML systems. This debt can lead to significant challenges in maintaining and evolving ML models, ultimately affecting their performance, reliability, and accuracy.

**Sources of Test Debt in ML**

*Training Data*

Inconsistent or poor-quality training data can lead to models that perform well during development but fail in production environments due to overfitting or data drift.

Bugs in the training data, such as noise, biased labels, or skew between training and test data, can introduce hidden faults that degrade model performance over time.

*Model Configuration and Hyperparameters*

Incorrect configuration of model parameters, such as choosing inappropriate activation functions or optimization algorithms, can introduce instability and reduce model effectiveness.

Ensuring the robustness of these configurations through continuous monitoring and validation is crucial to avoid accumulating test debt.

*Testing and Monitoring Procedures*

Insufficient or infrequent testing of ML models during development and deployment can lead to undetected faults and performance degradation.

Implementing comprehensive testing strategies, including property-based and metamorphic testing, helps identify and mitigate potential issues early in the development cycle.

*Execution Environment*

Differences between training and production environments can cause models to behave unpredictably, necessitating thorough testing across various environments to ensure consistency.

Regularization and fine-grained validation techniques, such as TFCheck, are recommended to maintain model stability and performance.

*Third-party Tools and Frameworks*

Dependency on third-party libraries and frameworks introduces an additional layer of complexity and potential sources of faults.

Rigorous validation of these dependencies and their integration into the ML pipeline is essential to prevent cascading failures and test debt.

**Mitigation Strategies**

*Comprehensive Testing Frameworks*

Developing and employing robust testing frameworks tailored for ML applications can significantly reduce test debt. These frameworks should support a variety of testing techniques, including unit tests, integration tests, and end-to-end tests.
Utilizing specialized tools like TFCheck and metamorphic testing frameworks can enhance the detection of subtle bugs and model inconsistencies.
*Continuous Monitoring and Validation*
Implementing continuous integration and continuous deployment (CI/CD) pipelines with built-in monitoring and validation checks helps maintain model integrity throughout its lifecycle.
Regular performance audits and regression testing ensure that new changes do not introduce unforeseen issues, thereby controlling test debt accumulation.
*Data Quality Assurance*
Establishing stringent data validation protocols to ensure the completeness, accuracy, and representativeness of training and test data is critical.
Automated tools for data cleaning and preprocessing can help identify and rectify data-related issues early, reducing the risk of introducing test debt.

**Conclusion**
Test debt in ML systems is a multifaceted issue arising from various sources, including data quality, model configuration, testing practices, and execution environments. Addressing test debt requires a comprehensive approach involving robust testing frameworks, continuous monitoring, and stringent data quality assurance. By implementing these strategies, organizations can mitigate the impact of test debt, ensuring the long-term reliability and performance of their ML models.

#### A.9.1.18 Versioning

Versioning Debt in machine learning (ML) and software engineering refers to the accumulated complexity and technical challenges arising from the management and evolution of various versions of software artifacts such as code, datasets, models, and configurations. This debt is akin to technical debt in traditional software but is exacerbated by the unique requirements and dependencies of ML systems.

**Key aspects include**
*Multiple Dependencies*: ML pipelines often consist of several interdependent stages, each with its own versioning requirements. Changes in any stage can necessitate changes in downstream stages, leading to complex dependency management.
*Data Dependencies*: Unlike traditional software, ML systems have strong dependencies on data versions. Changes in data schemas or the nature of the data itself can significantly impact model performance, requiring meticulous version control and documentation of data transformations and lineage.
*Semantic Versioning*: Semantic versioning is used to communicate the nature of changes in pipeline stages. It uses a MAJOR.MINOR.PATCH format to indicate backward-incompatible changes, backward-compatible additions, and backward-compatible bug fixes, respectively. However, this manual versioning can be error-prone and is often challenging to maintain consistently across different pipeline stages.
*Experimentation and Reusability*: Frequent experimentation in ML leads to multiple versions of datasets and models. Maintaining the ability to reproduce experiments while ensuring efficient reuse of existing computations is critical. This often involves creating derivations of datasets and ensuring that changes in configurations or code are tracked meticulously.
*Technical Debt Accumulation*: As ML systems evolve, the supporting code and infrastructure often accumulate significant technical debt. This includes the "plumbing and glue code" necessary to connect various ML libraries and packages. Over time, managing this debt requires substantial effort, especially in maintaining compatibility and performance across different versions of the system .
*Automation and Tooling*: Effective versioning requires robust tooling to automate many of the manual processes involved. Tools that can automatically track changes, manage dataset derivations, and ensure compatibility between different versions are crucial for reducing the burden of versioning debt.

### Conclusion

In summary, versioning debt in ML systems is a multifaceted problem that requires careful management of code, data, and configurations. Addressing this debt involves adopting systematic versioning practices, automating version control processes, and ensuring clear communication of changes across different components of the ML pipeline.

## A.10 Discussion

The scoping review identified various key types of technical debt specific to AI and machine learning (ML) systems. Each type presents unique challenges and impacts on system performance, maintainability, and reliability.

**Algorithm Debt** arises from suboptimal algorithm implementations, particularly in performance-critical AI frameworks. Its prevalence in deep learning frameworks highlights the need for continuous optimization to mitigate its impact on economic and safety-critical applications.

**Architectural Debt** is common in AI systems due to their rapid evolution and integration challenges. Issues like jumbled model architectures and pipeline jungles necessitate regular refactoring and adherence to best practices to maintain system scalability and reliability.

**Build Debt** accumulates from expedient but suboptimal development decisions. Effective management includes regular refactoring, enhanced testing, and comprehensive documentation. Intelligent techniques can further support debt management by identifying and prioritizing technical debt instances.

**Code Debt** in ML systems is characterized by issues such as code duplication, poor modularization, and dead experimental code paths. Strategies like refactoring, static code analysis, and adherence to best practices can mitigate these issues.

**Configuration Debt** stems from the complexities of managing numerous configurable options in ML systems. Addressing this debt involves improving the ease of configuration modifications, minimizing manual errors, and automating verification processes.

**Data Debt** is a critical concern due to AI's heavy reliance on data. It encompasses issues like poor data quality, unmanaged dependencies, and data relevance. Robust data management practices and continuous monitoring are essential to address data debt.

**Defect Debt** in ML frameworks arises from non-deterministic issues and inter-team dependencies, complicating defect management. Effective strategies include continuous monitoring and inter-team coordination to prevent long-term quality degradation.

**Design Debt** results from suboptimal design decisions in AI systems, often due to ad-hoc problem-solving approaches. Utilizing design patterns can significantly mitigate design debt by promoting modularity and maintainability.

**Documentation Debt** impacts the comprehensibility and maintainability of AI systems. Implementing structured documentation practices and leveraging automated tools can help reduce documentation debt and improve system transparency.

**Ethics Debt** emerges when AI systems are developed without adequately addressing ethical considerations. Proactively incorporating ethical analysis and robust oversight mechanisms throughout the AI lifecycle can mitigate ethical debt.

**Infrastructure Debt** pertains to the complex infrastructure required for AI systems. Addressing this debt involves robust configuration management, clear APIs, and continuous monitoring to ensure system reliability.

**Model Debt** in ML systems reflects the complexity and maintenance challenges of ML models. Strategies like early detection, holistic design, and comprehensive testing can mitigate model debt and maintain system performance.

**People Debt**, or social debt, impacts software projects due to suboptimal community interactions and organizational dynamics. Formalizing socio-technical decisions and investing in training can help manage people debt.

**Process Debt** refers to inefficiencies within workflows. Regular process audits and prioritizing process improvements can mitigate process debt and enhance organizational performance.

**Requirements Debt** in ML systems arises from complex dependencies and data requirements. Meticulous requirements engineering and proactive management can reduce requirements debt.

**Self-Admitted Technical Debt** (SATD) involves developers explicitly acknowledging technical debt through comments. Regular reviews and automated tools can help manage SATD effectively.

**Test Debt** in ML systems stems from inadequate testing practices. Comprehensive testing frameworks, continuous monitoring, and data quality assurance are critical to managing test debt.

**Versioning Debt** involves the complexities of managing versions of code, data, and models. Systematic versioning practices and robust tooling can reduce versioning debt and ensure compatibility across system components.

## A.11 Conclusion

Technical debt in AI and ML systems presents significant challenges to system performance, maintainability, and reliability. Addressing these debts requires a multifaceted approach involving continuous optimization, robust documentation and testing practices, ethical considerations, and effective management of data, configuration, and infrastructure. By adopting best practices and leveraging intelligent techniques, organizations can mitigate technical debt and ensure the long-term sustainability of their AI systems. Further research into standardized detection and management methods for various types of technical debt is essential to enhance the overall quality and reliability of AI-driven applications.

## A.12 Limitations

In all phases of our research, we aimed to mitigate potential threats to validity common in this type of research. However, our research is subject to several limitations.

Internal Validity

To reduce potential bias due to subjective interpretation by researchers, we defined a comprehensive set of keywords related to technical debt in AI-based systems and competing platforms (RQ1). We selected five data sources for our search to broaden the results, focusing primarily on peer-reviewed papers. We also performed snowballing to ensure the inclusion of papers that may have been missed in the original search.

In the first selection phase, we reviewed the title, abstract, and keywords of the obtained results. The first author then read each article in its entirety, making detailed comments on the content, key points, and findings relevant to our research. The second author reviewed these comments and assessed the quality of the findings. Any papers in doubt were discussed jointly, ensuring consensus on their suitability for inclusion in the final results. A similar procedure was followed for data extraction and analysis.

Despite our adherence to a rigorous protocol, the inherent subjectivity in the selection process means that other researchers using our strategy may obtain slightly different results.

External Validity

The field of technical debt in AI-based systems is constantly evolving, with new categories emerging and existing ones being modified. Issues previously highlighted are often re-examined from different perspectives or within different application frameworks, revealing new dimensions. Consequently, nuanced findings may emerge regarding the definition, scope, and impact of technical debt in AI-based systems as these systems mature and see increased adoption in software engineering. This is particularly true in unexplored areas such as AI-based competition platforms, where evolving insights are to be expected.

## A.13 Methods

To ensure a comprehensive and systematic identification of relevant studies, we followed a rigorous multi-phase study selection process as outlined by the PRISMA (Preferred Reporting Items for Systematic Reviews and Meta-Analyses) guidelines. The PRISMA flow diagram below illustrates each step of this process, including the number of records identified, screened, assessed for eligibility, and ultimately included in the scoping review.

**Identification**: We conducted an extensive search across multiple databases (Google Scholar, ACM Digital Library, IEEE Xplore, Scopus, and Springer) using a predefined set of keywords related to technical debt in AI-based systems. Additional records were identified through manual searches and snowballing techniques to capture relevant studies that might have been missed in the initial search.

**Screening**: After removing duplicates, the titles and abstracts of the remaining records were screened for relevance based on predefined inclusion and exclusion criteria. This initial screening aimed to filter out studies that were clearly not pertinent to our research objectives.

**Eligibility**: Full-text articles of potentially relevant studies were retrieved and assessed for eligibility. This detailed evaluation ensured that only studies meeting all inclusion criteria were considered for the final analysis.

**Included**: The final set of studies included in the scoping review was determined through consensus among the researchers, ensuring a comprehensive and unbiased selection of relevant literature.

The flow diagram provides a visual summary of this process, highlighting the number of records at each stage and the reasons for exclusion where applicable.

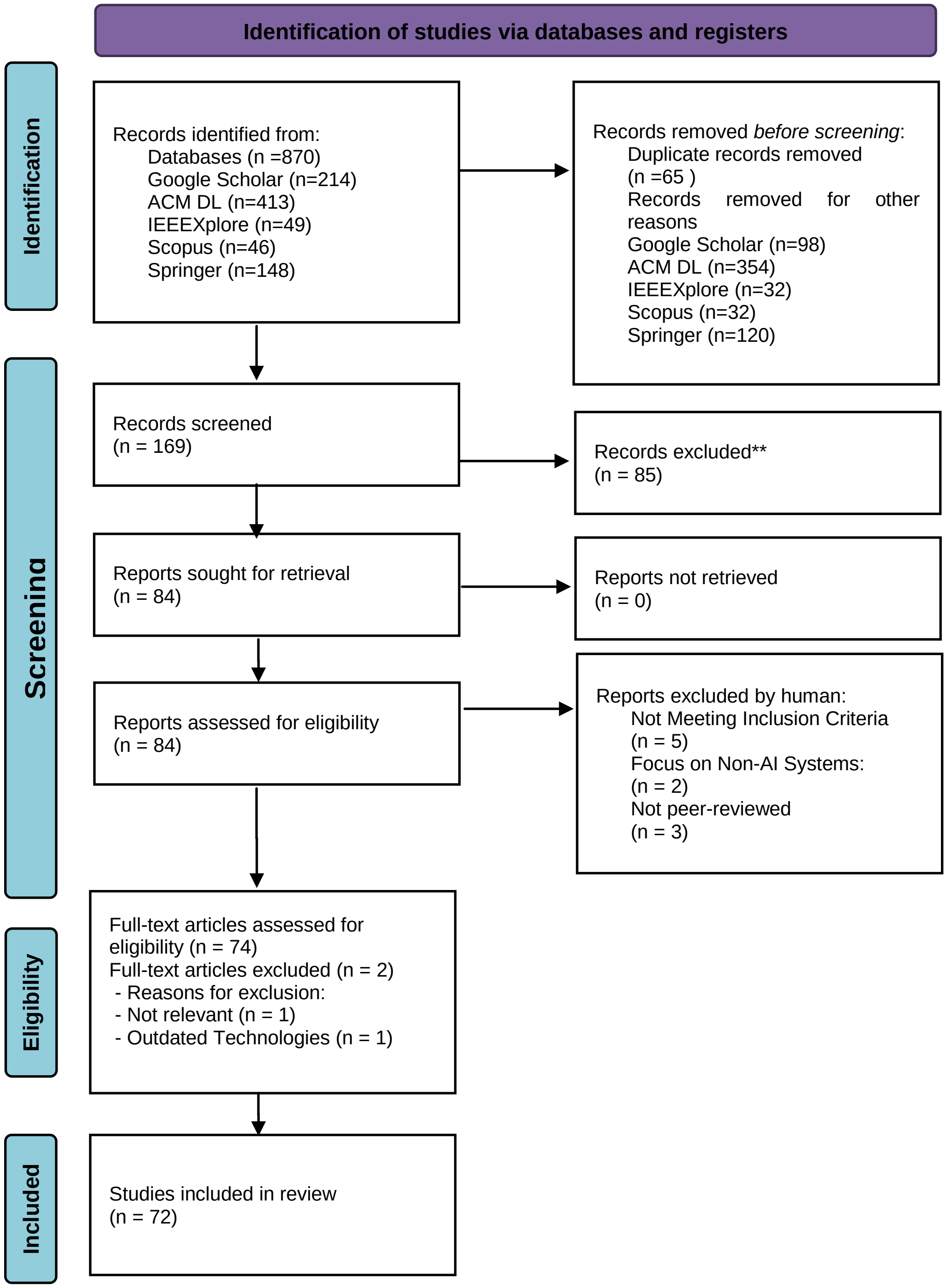

In summary, the study selection process started with 870 records identified through database searching. After removing duplicates and records for other reasons, 169 records were screened. From these, 85 were excluded, and 84 full-text articles were sought and successfully retrieved for eligibility assessment. Of these, 10 were excluded for various reasons, leaving 74 full-text articles, of which 2 were further excluded. Ultimately, 72 studies were included in the qualitative synthesis, ensuring a comprehensive and rigorous review of the literature on technical debt in AI-based systems.

## APPENDIX B

### B.1 Accessibility Debt

*Definition:* Accessibility Debt refers to the barriers that participants encounter due to the lack of immediate usability of platform technologies.
*Problem:* The primary issue with Accessibility Debt is that it can dissuade potential participants from engaging with AI/ML competition platforms. This is due to the challenges they face in utilizing the platform's components quickly and efficiently. If these barriers are not addressed, the competitive value and attractiveness of such platforms diminish, leading to decreased participation and devalued competitions.
*Example:* Consider an AI-based competition platform where participants must navigate a complex setup process before they can begin working on their assigned tasks. If the process involves multiple steps, unclear instructions, or requires significant troubleshooting, participants may become frustrated and disengaged. This scenario exemplifies Accessibility Debt, as the immediate *usability* of the platform is compromised, hindering participant involvement and satisfaction.
*Stakeholder:* Organizer: They are responsible for ensuring that the platform is accessible and user-*friendly*, providing necessary tools and resources to participants efficiently. By addressing Accessibility Debt, organizers can enhance participant engagement and the overall success of the competitions.

### B.2 Algorithm debt

*Definition:* Algorithm debt refers to technical debt arising from the choice, implementation, and maintenance of algorithms within AI systems, excluding issues directly related to the model's structure or data handling. This debt encompasses challenges associated with selecting appropriate algorithms, optimizing their performance, and ensuring they remain suitable as the system scales or as requirements evolves.
*Problem:* The problem of algorithm debt in AI-based systems often stems from the use of algorithms that are either too simplistic or overly complex for the task at hand, poorly optimized, or inefficient in terms of computational resources. Such choices may initially simplify development but can lead to increased costs and reduced system performance in the long run. For instance, an algorithm that is not scalable might handle initial data volumes well but becomes a bottleneck as data grows, requiring costly re-engineering efforts.
*Example:* In the context of AI-based competition platforms, an example of algorithm debt could occur when an algorithm designed for matchmaking in online games does not effectively adapt to changes in player skills and behaviors over time. Initially, the algorithm may work well, but as the variety of players increases, its inability to adapt could result in poor matchmaking, increased wait times, and player dissatisfaction. The platform may then face significant technical debt as the original algorithm requires substantial modification or replacement to meet the evolved needs of its user base.
This example highlights how crucial it is to anticipate the long-term needs of the system when choosing algorithms, and to plan for their evolution as part of the platform's ongoing development strategy to avoid accruing algorithm debt. This example illustrates how the need for rapid development in competitive AI platforms can lead to significant algorithm debt, impacting the platform's long-term capability to perform reliably and accurately.

### B.3 Architectural debt

*Definition:* Architectural debt refers to the complexities arising from the intricate integration of architectural components with their foundational data within AI-based systems.
*Problem:* This type of debt can precipitate intricate and indeterminate dependencies between system components and their corresponding datasets. It often results in architectures that are challenging to evaluate due to their complex compositions. Moreover, it can lead to the emergence of undeclared entities that utilize AI models, further complicating the governance and maintenance of these systems.
*Example:* At a global AI-based game development competition, participants faced significant challenges due to architectural debt within their gaming platforms. The competition's objective was to develop AI-driven games that could adapt to user behaviors dynamically. However, the deeply entangled architectures of these systems, combining multiple data sources and AI components, led to unpredictable game behaviors. For instance, changes in player data inputs drastically altered AI responses, not due to algorithmic intent but because of hidden dependencies within the game’s architecture. Additionally, some components of the games began consuming additional data without clear documentation, leading to further inconsistencies and performance issues. These problems made it difficult for developers to assess or modify their games effectively, ultimately affecting the overall competition experience and the performance of the AI models in a live environment. This scenario highlights how architectural debt can compromise the functionality and innovation in AI-driven gaming competitions.

### B.4 Build debt

*Definition:* Build debt encompasses the suboptimal dependencies within and between AI models, both internal and external. This refers to the reliance on inefficient or obsolete components and processes in the development and deployment of AI systems.
*Problem:* This type of debt becomes apparent through the implementation of inefficient building processes and the continued use of outdated dependencies. These practices can lead to reduced system efficiency and performance, increased maintenance costs, and hindered scalability. Ultimately, build debt can compromise the robustness and responsiveness of AI systems, affecting their ability to adapt to new requirements or integrate with modern technologies.

*Example:* In an AI-based game competition, a team's performance was significantly hampered by build debt. Their AI model relied on several outdated machine learning libraries and deprecated data processing frameworks, which had not been updated due to build debt. During the competition, this resulted in slower response times and subpar decision-making by the AI, particularly when compared to competitors using more modern and efficient technologies. The team's reliance on these outdated dependencies not only diminished the AI's effectiveness but also showcased the critical need for continuous updates and integration of current technologies to maintain competitiveness in rapidly evolving environments.

## B.5 Code debt

*Definition:* Code debt characterizes the less-than-ideal coding practices frequently observed in the development of AI models, which often stem from the experimental nature inherent to AI research and development.

*Problem:* This type of debt is evident in several critical areas. For instance, AI-based systems may contain residual experimental code that is ineffectively integrated into the production environment. Moreover, the transition from experimental models to fully deployable software often lacks optimal refactoring, leading to inefficiencies. The algorithmic complexity of AI models can also introduce additional code deficiencies that undermine the system's robustness and efficiency.

*Example:* An AI game competition was organized where different AI systems competed in a classic game of chess. Each AI participant was designed to think several moves ahead and adapt to their opponent's strategies. The competition served to highlight the strengths and weaknesses of each AI's decision-making abilities. It provided a straightforward environment to observe how well each AI could handle complex decision trees and react to unexpected moves. This event also helped developers identify areas where their AI models could be improved, especially focusing on minimizing code debt by refining the algorithms for better performance and reliability in real-time decisions.

## B.6 Configuration debt

*Definition:* Configuration debt encapsulates the deficiencies prevalent within the configuration frameworks of AI-based systems. This term specifically refers to the complexities and shortcomings of configuration processes, including the use of convoluted, insufficiently documented, unversioned, or untested configuration files.

*Problem*: Configuration debt introduces significant vulnerabilities to machine learning systems by fostering errors in configuration that can deplete valuable time and computational resources and cause delays in production. This debt hampers the ability to accurately modify and understand configurations, complicating the evaluation of the effects of changes and the comparison of configurations across different iterations. Moreover, poor management of configuration settings intensifies these issues by obstructing the precise specification, tracking, and reproducibility of configuration alterations. This results in difficulties in replicating experiments and ensuring system reliability. Effective management of configuration settings is crucial to alleviate these problems and enhance the efficiency, reproducibility, maintainability, and transparency of AI-based systems.

*Example:* During a high-profile AI-based game competition, one team struggled with significant configuration debt. Their gaming AI had multiple configuration files that were complex and poorly documented, making it difficult for new team members to understand and modify the settings efficiently. As the competition progressed, the need to adapt to opponents' strategies became critical. However, due to unversioned and untested configurations, changes made under pressure led to errors that were not detected until too late. This resulted in the AI performing unpredictably, costing the team crucial matches and demonstrating the impact of configuration debt on competitiveness and system reliability.

## B.7 Data debt

*Definition:* Data debt pertains to the shortcomings in the collection, management, and application of data within AI-based systems, characterized by issues such as poor data quality, unregulated data dependencies, and inadequate data relevance.

*Problem:* These deficiencies can compromise the efficiency and precision of AI models, posing challenges to the system's reliability and future adaptability. Data debt introduces risks that may impede the ongoing development and operational effectiveness of AI systems, potentially leading to erroneous outputs and strategic misalignments in long-term system evolution.

*Example:* In a high-profile AI-based gaming competition, a team encountered significant setbacks due to data debt in their system. Their AI model, designed to adapt strategies dynamically against opponents, suffered from the use of outdated and poorly curated datasets, leading to issues with data relevance and quality. This compromised the model’s ability to make accurate decisions during gameplay, often choosing less effective strategies or failing to anticipate opponent moves. The problem was exacerbated by unmanaged data dependencies that weren't identified until the competition, where the need for real-time data processing highlighted these critical shortcomings. These issues not only diminished the AI's performance in a competitive setting but also raised concerns about the long-term viability and scalability of the system, underscoring the crucial impact of maintaining rigorous data management practices to support AI effectiveness in dynamic environments.

## B.8 Defect debt

*Definition:* Defect debt describes the phenomenon where code exhibits unexpected behavior, compelling developers to defer its rectification owing to the complexity of the repairs or constraints on time [7].

*Problem:* This type of debt embodies unresolved anomalies or bugs within a software system. Encountering code that does not perform as anticipated, developers might opt to postpone necessary fixes due to the repair's intricacy or pressing deadlines. Such decisions contribute to the accumulation of technical debt, characterized by persisting known issues which may lead to further bugs and erratic software behavior. Over time, if the focus remains on adding new features rather than addressing existing defects, or if resolution efforts are hindered by resource limitations, defect debt may escalate significantly.
*Example:* During an AI-based game competition, a team encountered significant challenges due to defect debt. Their AI system, crucial for decision-making in the game, had a known bug that caused it to misinterpret game rules under specific conditions. Despite awareness of the issue, the team had previously postponed fixing the bug due to its complexity and the looming competition deadline. As a result, during a critical match, the AI behaved unpredictably, leading to a crucial loss. This incident highlighted the risks of accumulating defect debt, especially when enhancements are prioritized over essential maintenance and bug resolution in competitive environments.

## B.9 Design debt

*Definition:* Design debt refers to suboptimal elements within the architecture of artificial intelligence systems, which can complicate their maintenance and future evolution.
*Problem:* Design debt leads to systemic challenges that reduce modularity and complicate codebase maintenance, increasing complexity and reducing component reusability. Integration issues cause data inconsistencies, communication failures, and coordination problems. This debt also deters the adoption of new technologies, maintaining outdated methods that degrade system performance and scalability. Furthermore, it adds rigidity, reducing adaptability to new requirements or innovative features and raising costs for system modifications. Design debt also limits scalability, impairing the system's ability to handle increased data or user loads, thus degrading performance and efficiency. Addressing design debt is essential to improve system modularity, integration, technological adoption, flexibility, maintainability, and scalability, ensuring sustained robust performance and adaptability.
*Example:* In an AI-based game competition, design debt becomes prominently problematic when systems fail to adapt to dynamic gaming environments. For example, a gaming AI designed with rigid algorithms may excel in initial, simpler scenarios but struggles as complexities increase. Design debt in the AI's architecture – such as hard-coded strategies or poor data integration – prevents efficient adaptation to new game rules or strategies introduced mid-competition. This leads to diminished performance, where the AI cannot scale its capabilities or incorporate innovative tactics, ultimately reducing its competitive edge and increasing the time and resources required for essential updates or enhancements.

## B.10 Documentation debt

*Definition:* Documentation debt refers to the gaps or inadequacies in the documentation associated with AI-based systems. This includes the lack of comprehensive documentation covering system features and the assumptions underlying the data utilized in these systems.
*Problem:* The absence or insufficiency of detailed documentation complicates the understanding and maintenance of AI systems over time. This deficit can obstruct effective knowledge transfer, hinder system scalability, and increase the likelihood of errors during updates or when integrating new components, ultimately impacting the long-term sustainability and operability of the system.
*Example:* During an AI-based game competition, teams noticed issues stemming from documentation debt. The AI systems deployed lacked detailed documentation on their decision-making processes and data assumptions. This led to confusion among the developers when their AIs behaved unpredictably or failed to adapt to game dynamics effectively. Teams struggled to diagnose problems or make timely adjustments, as the necessary information was either missing or unclear. This situation underscored the critical need for thorough documentation to ensure that AI systems are fully understood and can be efficiently managed and improved over time, especially in competitive scenarios.

## B.11 Ethics debt

*Definition:* Ethics debt refers to the shortcomings concerning the ethical dimensions of AI-based systems, including issues such as algorithmic fairness, prevalent prediction biases, and a lack of transparency and accountability.
*Problem:* The realm of AI ethics encounters multiple challenges that stem from a diminished influence over decision-making processes, insufficient enforcement mechanisms, and an over-reliance on ethical guidelines rather than binding legal standards. This often results in the neglect of broader social contexts and relationships, contributing to a lack of diversity within the AI community and the exclusion of vital ethical considerations. The ramifications of ethics debt are manifold, encompassing unintended harmful effects and the malicious use of AI technologies. Such consequences include job displacement, unsupervised experimental applications, and significant data breaches. Moreover, there exists a substantial risk of developing biased algorithms and unsafe products, precipitously deploying immature applications, and the potential exploitation of AI technologies by malicious entities, such as criminal hackers.
*Example:* In an AI-based game competition, a team deployed an AI model designed to predict and counter opponents' moves. Unfortunately, the model exhibited significant ethics debt due to overlooked issues like algorithmic fairness and transparency. The model, trained predominantly on data from games played by a non-diverse group, failed to fairly assess strategies used by a wider range of competitors. This bias led to inaccurate predictions and strategic blunders, compromising the team's performance. The incident highlighted the consequences of neglecting ethical considerations in AI development, demonstrating how ethics debt can undermine fairness and effectiveness in competitive environments.

## B.12 Infrastructure debt

*Definition:* Infrastructure debt encapsulates the inadequacies inherent in the implementation and operational management of artificial intelligence (AI) pipelines and models.

*Problem:* This type of debt introduces significant operational and reproducibility challenges within AI systems. It frequently manifests as overly complex infrastructures that integrate multiple AI pipelines, leading to suboptimal resource distribution for the training and testing of AI models. Additionally, it exacerbates the difficulty in effectively monitoring and debugging AI systems, thereby compromising their reliability and performance.

*Example:* During a high-profile AI-based game competition, infrastructure debt became evident as teams struggled with the systems set up for training and testing their game-playing models. The competition's infrastructure was initially designed to handle multiple concurrent AI pipelines efficiently. However, as the competition progressed, it became clear that there were significant deficiencies in resource allocation and system integration. Some teams experienced delays in model training due to limited GPU resources, while others faced challenges in consistently reproducing game strategies due to varying system performances. Additionally, the lack of robust monitoring and debugging tools meant that identifying and resolving these issues was time-consuming, leading to uneven playing fields and questioning the fairness and integrity of the competition. This scenario highlights how infrastructure debt can severely impact the operational success of AI-driven initiatives in competitive environments.

## B.13 Model debt

*Definition* Model debt refers to the accumulation of suboptimal practices within the lifecycle of artificial intelligence models, encompassing the design, training, and management phases. This encompasses issues related to inadequate feature selection, improper tuning of hyperparameters, and ineffective strategies for model deployment.

*Problem:* Model debt manifests through several challenges, such as poorly chosen features, neglected adjustments of hyperparameters, and deficient deployment architectures. Such deficits may complicate the maintenance of models and diminish their predictive accuracy, thereby undermining the system's overall effectiveness and reliability.

*Example:* In an AI-based game competition, teams leverage complex models to predict opponent strategies and optimize gameplay. One team, however, experienced significant model debt. Initially, their AI model showed promise, but over time, it became apparent that the feature selection was too narrow, focusing solely on short-term gains rather than strategic depth. Furthermore, hyperparameters were not adequately tuned for the dynamic nature of the game, and the deployment strategy failed to adapt to evolving game scenarios. This resulted in the model underperforming during critical matches, as it could not accurately predict or counter diverse strategies employed by competitors.

## B.14 People debt

*Definition:* People debt signifies the shortcomings associated with the expertise and capabilities of individuals engaged in the development and maintenance of AI-based systems. This category also encompasses the interpersonal and interdepartmental challenges, particularly between data scientists and software engineers, which can hinder the effective collaboration necessary for constructing robust AI-enabled systems.

*Problem:* Such deficiencies may impede the comprehensive understanding and adept handling of AI technologies among team members, fostering suboptimal development practices. Notably, this form of debt manifests as ambiguous role delineations, misaligned practices and expectations, communication barriers, and clashes over technical competencies, all of which can significantly disrupt project progress and quality.

*Example:* In an AI-based game competition, a team encountered significant challenges due to people debt. The team, comprised of data scientists and software engineers, struggled with unclear roles and poor communication, which led to inefficiencies in integrating AI strategies into the gaming software. As the competition progressed, these issues became apparent when their AI consistently misinterpreted game rules or reacted inappropriately to opponent moves. The lack of cohesive expertise and collaboration not only reduced the AI's competitiveness but also highlighted the critical impact of team dynamics on the development and performance of AI systems in high-stakes environments.

## B.15 Process debt

*Definition:* Process debt in AI-based systems, as an extension of general software systems, includes inefficiencies in development, deployment, or maintenance, covering data collection, preprocessing, training, evaluation, deployment, monitoring, and iteration.

*Problem:* Arising from suboptimal practices or shortcuts in the AI development lifecycle, such as inadequate data preprocessing, poor model evaluation, manual error-prone deployment, insufficient monitoring, and delayed updates, process debt compromises model accuracy, amplifies technical debt, inflates costs, and complicates maintenance and evolution of AI systems.

*Example:* In an AI-based game competition, process debt became evident when developers, pressed for time, opted for rapid yet suboptimal methods. The game, designed to evolve based on player interactions, initially performed well. However, the team skipped critical steps in data preprocessing and deployed the AI models manually without thorough testing. As the competition progressed, the lack of robust evaluation and monitoring meant that the AI did not adapt effectively to player strategies or update with new data. This resulted in increasingly predictable and exploitable AI behavior, reducing the game's challenge and appeal. The initial shortcuts led to higher operational costs and technical challenges, as the team struggled to refine and maintain the system. The process debt accumulated not only detracted from the player experience but also constrained the game's ability to evolve, demanding significant rework to address these foundational issues.

## B.16 Requirements debt

*Definition:* Requirements debt pertains to the inadequacies in the articulation and management of system requirements essential for the effective development and progressive refinement of AI-based systems.

*Problem:* This type of debt emerges when system requirements are ambiguous or insufficiently specified, complicating the accurate identification and fulfillment of stakeholder expectations. Such deficiencies can hinder the development process, leading to a system design and implementation that may not fully align with the intended functionalities or performance criteria, thereby affecting the overall efficacy and utility of the system.

*Example:* In a recent AI-based game competition, a team experienced significant setbacks due to requirements debt. Initially, the AI system was designed without a clear understanding of the competition's complex scoring rules and diverse gameplay scenarios. As a result, the AI struggled to adapt its strategies effectively during the matches. This misalignment between the AI's capabilities and the actual game requirements led to underperformance against competitors whose systems were better tailored to the specific demands of the competition. This example illustrates how poorly defined requirements can severely limit the effectiveness and adaptability of AI systems in dynamic environments.

## B.17 Self-Admitted Technical Debt (SATD)

*Definition:* Self-Admitted Technical Debt (SATD) comprises design and implementation choices recognized by developers as suboptimal during the software development process. These decisions are explicitly acknowledged within code comments, which identify aspects of the project needing further refinement or completion in the future. [XX].

*Problem:* SATD represents a specific category of technical debt that is both visible and quantifiable, distinctly articulated through annotations in the source code. This type of debt includes substandard development practices, incomplete documentation, acknowledged bugs, and deficiencies in code, testing, or software requirements. If not promptly addressed, SATD can substantially degrade the quality and maintainability of software, emphasizing its critical role in the development lifecycle.

*Example:* During an international AI-based game competition, a development team used an AI model that was laden with Self-Admitted Technical Debt (SATD). The developers had noted in the code comments that certain algorithms were hastily implemented and needed optimization. As the competition progressed, the AI model began to exhibit erratic behaviors, particularly in handling unexpected strategies from opponents. This was primarily due to the acknowledged but unaddressed inefficiencies in the model's decision-making algorithms. The team's failure to resolve these issues before the competition led to poor performance and demonstrated how SATD could critically undermine the functionality and success of AI systems in real-time competitive scenarios.

## B.18 Test debt

*Definition:* Test debt encompasses the shortcomings in testing practices within AI systems, particularly highlighting the insufficient testing of AI models and pipelines, as well as a deficiency in advanced testing methodologies necessary for evaluating data quality and distribution.

*Problem:* The probabilistic characteristics of certain AI algorithms add a layer of complexity to the testing processes. These complexities make it especially challenging to ensure comprehensive testing coverage and reliability, thereby increasing the risk of undetected issues in AI system behavior under varied operational conditions. This type of debt may lead to less predictable and potentially unreliable AI system performance.

*Example:* In an AI-based game competition, various AI models were pitted against each other in a strategy game. However, due to existing test debt, the competition faced challenges. The AIs had not been adequately tested for their ability to handle the stochastic elements of the game, such as random events and unpredictable opponent strategies. This lack of rigorous testing resulted in some AI models performing inconsistently, displaying erratic behavior, or failing to adapt to new scenarios presented during the competition. The event highlighted critical gaps in testing practices, emphasizing the need for more robust testing frameworks to ensure AI reliability and effectiveness in dynamic environments.

## B.19 Versioning debt

*Definition:* Versioning debt refers to the substandard practices associated with the versioning of AI models and their corresponding training and testing datasets. This includes insufficient version control systems or the absence of robust versioning mechanisms.

*Problem:* Such inadequacies complicate the tracking and management of different iterations of models and datasets, which is essential for the orderly evolution of AI systems. Furthermore, the lack of effective versioning undermines the reproducibility of the system, presenting challenges in validating outcomes and replicating previous states, thereby affecting the reliability and scientific integrity of the AI system development process.

*Example:* During an AI-based game competition, one team faced challenges stemming from versioning debt. They had multiple versions of their AI model, each trained with slightly different datasets, but poor version control practices made it difficult to identify which version was most current or effective. This confusion led to deploying an outdated model that underperformed during the competition. The inability to track and manage model versions not only hindered the team's performance but also showcased the critical importance of robust versioning systems in maintaining the reliability and competitiveness of AI systems in dynamic and competitive settings.

# APPENDIX C

## C.1 Algorithm debt

***Question1****:* Have you checked if the framework you are using has technical debt or may introduce glitches or incompatibility in your application?
*Stakeholder*: Organizers & Participants
Score: 3
*Justification:* Identifying and understanding the technical debt within the framework is essential. It can affect the application's performance, scalability, and even the user experience. Glitches and incompatibilities can lead to a poor reputation and user frustration.
*Example:* If the platform is using an outdated version of TensorFlow, it might miss out on new optimizations that could speed up model training. If the chosen framework has a history of memory leaks, it could affect the platform's ability to scale and handle multiple concurrent games.

## C.2 Architecture – Design debt

***Question2****:* Have you effectively separated concerns and ensured that code reuse does not lead to tightly coupled components?
*Stakeholder*: Organizers
Score: 5
*Justification:* Poor separation of concerns can lead to a tangled system that is hard to debug and evolve, significantly increasing technical debt.
*Example:* Using interfaces or abstract classes to define contracts between components, so they can be easily swapped or modified without affecting others.
***Question3****:* Pipeline Jungle - Is it possible to maintain a single controllable, straightforward pipeline of ML components?
*Stakeholder*: Organizers & Participants
Score: 3
*Justification:* A convoluted pipeline can be difficult to maintain and upgrade, contributing to technical debt over time.
*Example:* Multiple data preprocessing steps scattered across the pipeline, making it hard to track changes and fix bugs.
e.g. Have you designed separate modules/services for data collection and data preparation?
e.g Have you checked for improper reuse of complete AI components or pipelines?
***Question4****:* Does your system include glue code?
*Stakeholder*: Organizers & Participants
Score: 2
*Justification:* Glue code is often a quick fix that becomes permanent, increasing technical debt as the system scales.
*Example:* Temporary scripts that become a permanent part of the workflow, complicating future updates.
***Question5****:* Have you avoided reusing a slightly modified complete model (correction cascades)?
*Stakeholder*: Participants
Score: 3
*Justification:* Correction cascades can create a maintenance nightmare, adding to the technical debt each time the base model is updated.
*Example:* A small change in the base model requiring adjustments in all derived models
***Question6****:* Have you designed the environment for prototyping ML models to prevent the need to re-implement from scratch for production?
*Stakeholder*: Organizers
Score: 4
*Justification:* The need to re-implement models for production can lead to redundant work and increased technical debt.
*Example:* A model developed in a research setting that requires significant refactoring to be deployed in production.

## C.3 Build debt

***Question7****:* Have you installed a proper version control system for model, training and test data?
*Stakeholder*: Organizers & Participants
Score: 4
*Justification:* Ensuring that your app is free from bad or suboptimal dependencies is crucial for maintaining the integrity, performance, and security of your application. Bad or suboptimal dependencies can contribute to build debt by causing conflicts, reducing performance, and making the build process more fragile.
*Example:* A reinforcement learning platform might rely on a specific machine learning library for neural network computations. If this library is not kept up-to-date or is known to have performance issues, it could hinder the platform's ability to scale or adapt to new challenges.
***Question8****:* Have you installed a proper version control system for model, training and test data?
*Stakeholder*: Organizers
Score: 3
*Justification:* Cross-platform compatibility is important for reaching a wider audience and ensuring that your app can operate on various devices and operating systems. Inability to build consistently across different platforms can be indicative of build debt, as it suggests a lack of portability and potential issues with platform-specific dependencies.
*Example:* Ensure that the platform can be deployed on both Windows and Linux systems, which might require different sets of dependencies and configurations.

## C.4 Code debt

***Question9****:* Have you identified and refactored low-quality, complex, and duplicated code sections, including dead codepaths and centralized scattered code, while ensuring clear component and code APIs?
*Stakeholder*: Organizers
Score: 5
*Justification:* Dead codepaths can significantly impact performance and maintainability of the codebase. Identifying and removing them ensures efficient resource utilization and reduces the chances of bugs or unexpected behavior. Low-quality code can impede readability, maintainability, and scalability. Detecting and addressing it early can prevent technical debt accumulation and streamline development efforts. Complex code can be challenging to understand, debug, and maintain. Identifying complex sections allows for simplification, leading to improved code quality and developer productivity. Centralizing scattered and duplicated code improves maintainability and reduces the likelihood of inconsistencies. It promotes code reuse and ensures uniformity across the codebase.

*Example:* Suppose a section of the platform's codebase contains duplicate functions for calculating player scores. Refactoring would involve removing the redundancy, centralizing the score calculation, and ensuring that the remaining function has a clear API that can be used by other components of the platform. An example of this type of technical debt would be a function that has grown too large and complex over time, making it difficult to understand and modify. Refactoring it into smaller, well-named functions can improve readability and maintainability. After identifying a set of functions with tightly coupled logic, you refactor them to reduce dependencies and define clear interfaces. This makes the code easier to understand and modify, which is beneficial when adding new features or debugging. An example here would be updating an API to use RESTful principles, making it easier for developers to understand how to interact with it and for the system to integrate with other service.

## C.5 Configuration debt

***Question10:*** Is it easy to specify a configuration as a small change from a previous configuration?
*Stakeholder*: Organizers
Score: 3
*Justification:* In RL, quick experimentation is crucial. Being able to specify configuration changes easily allows rapid iteration and model improvement.
*Example:* A platform allows incremental changes to the learning rate of an RL agent by modifying a single line in a YAML file, facilitating quick experimentation.
***Question11:*** Is it easy to specify a configuration as a small change from a previous configuration?
*Stakeholder*: Organizers
Score: 4
*Justification:* Proper documentation facilitates understanding and maintenance of configurations.
*Example:* A platform's configuration file lacks comments explaining the purpose of each parameter, leading to confusion among developers.
***Question12:*** Is it easy to specify a configuration as a small change from a previous configuration?
*Stakeholder*: Organizers
Score: 4
*Justification:* Reviewing configurations is crucial for identifying errors, optimizing performance, and maintaining consistency. Testing ensures that configurations perform as expected under different conditions. Rigorous testing ensures robustness and reliable performance. Review and testing are vital to catch and identify errors early, optimizing performance, maintaining consistency and ensure the system behaves as expected under various configurations.
*Example:* A platform undergoes a peer review process for configuration changes, followed by automated tests to validate the new settings.

## C.6 Data debt

***Question13:*** Have you checked for spurious data?
*Stakeholder*: Participants
Score: 4
*Justification:* Spurious data can introduce noise and lead to overfitting, which increases technical debt due to the need for additional debugging and retraining.
*Example:* Identifying and removing outliers from player score datasets that do not align with expected patterns.
***Question14:*** Have you checked your data for accuracy?
*Stakeholder*: Participants
Score: 5
*Justification:* Inaccurate data can mislead the training process, resulting in models that perform poorly in real-world scenarios, thus accumulating technical debt.
*Example:* Verifying the correctness of reward signals in the game environment to ensure they reflect the intended game mechanics.
***Question15:*** Have you checked your data for completeness?
*Stakeholder*: Participants
Score: 4
*Justification:* Incomplete data can result in underfitting and poor generalization, leading to technical debt when the model fails to perform as expected.
*Example:* Ensuring that the dataset includes a wide range of scenarios that a player might encounter in the game.
***Question16:*** Have you checked your data for trustworthiness?
*Stakeholder*: Participants
Score: 4
*Justification:* Untrustworthy data can stem from biased or manipulated sources, increasing technical debt by causing the model to learn incorrect behaviors.
*Example:* Assessing the reliability of data sources, such as player feedback, to confirm they are not influenced by external incentives.
***Question17:*** Have you performed testing on the input features?
*Stakeholder*: Participants
Score: 3
*Justification:* Testing input features is essential to ensure they are predictive and relevant, reducing technical debt by preventing the inclusion of irrelevant or redundant features.
*Example:* Conducting feature selection to determine the most significant inputs for predicting player engagement.
***Question18:*** Have you checked your data for data relevance
*Stakeholder*: Participants
Score: 3

*Justification:* Irrelevant data can lead to a model that does not adapt well to the task, increasing technical debt through unnecessary complexity and maintenance.
*Example:* Filtering out gameplay data that does not contribute to the learning objective, such as background art elements.

## C.7 Defect debt

***Question19:*** Have you checked that there is no error in the training data collection that would cause a significant training data set to be lost or delayed?
*Stakeholder*: Participants
Score: 5
*Justification:* Ensuring error-free training data collection is paramount. Errors in training data can introduce biases and inaccuracies that compound over time, leading to significant technical debt.
*Example:* For example, if a racing game's training data incorrectly labels off-track excursions as successful maneuvers, the model may learn to drive off-track, requiring extensive retraining and data cleansing later
***Question20:*** Have you made the right choice in the hyperparameter values?
*Stakeholder*: Participants
Score: 4
*Justification:* Choosing the right hyperparameters is essential for model performance and efficiency. Incorrect hyperparameter values can lead to suboptimal performance or slow convergence, impacting the overall effectiveness of the model.
*Example:* For instance, incorrect learning rates can cause a model to converge too slowly or not at all, impacting the speed of iteration and potentially leading to a backlog of updates.
***Question21:*** Have you made sure that there is no degradation in view prediction quality due to data changes, different code paths, etc.?
*Stakeholder*: Participants
Score: 4
*Justification:* Ensuring that changes in data or code paths do not degrade view prediction quality is critical in RL games. Even minor degradation in view prediction quality can affect the player's experience and the game's overall performance.
*Example:* An example is a strategy game where unit strengths may change over time; if the model cannot adapt to these changes, its strategies may become obsolete.
***Question22:*** Have you quality inspected and validated the model for adequacy before releasing it to production?
*Stakeholder*: Participants
Score: 5
*Justification:* Quality inspection and validation of the model before releasing it to production are essential in RL games to ensure that the model performs adequately and meets the desired performance criteria. Releasing an inadequately validated model can lead to poor player experience which is a form of technical debt that is often expensive to address post-release and potentially damage the game's reputation.
*Example:* For example, a model that hasn't been validated for a shooter game might misclassify in-game objects, leading to frustrating gameplay and the need for urgent patches.
***Question23:*** Have you implemented mechanisms for rapid adaptation and regular updates to maintain the model's efficiency and relevance in response to changes in data, features, modeling, or infrastructure?
*Stakeholder*: Organizers & Participants
Score: 4
*Justification:* Implementing mechanisms for rapid adaptation is essential in the fast-paced environment of games, where data and features can change frequently. Without this, the model may quickly become outdated, accumulating technical debt.
*Example:* For example, in a multiplayer online battle arena (MOBA) game, new characters and abilities are introduced regularly; without rapid adaptation, the model's strategies could become ineffective.

## C.8 Documentation debt

***Question24:*** Is Requirement Documentation available?
*Stakeholder*: Organizers
Score: 5
*Justification:* Requirement documentation is crucial for understanding the goals and objectives of the project, as well as the functionalities expected from the system. Without it, development might lack direction or focus, leading to potential misunderstandings and misalignments.
*Example:* If the platform's matchmaking algorithm requirements are not well-documented, developers might implement incorrect or suboptimal features that could require significant rework.
***Question25:*** Is Technical Documentation available?
*Stakeholder*: Organizers
Score: 5
*Justification:* Technical documentation provides insight into the system's architecture, components, and functionalities from a technical perspective. It aids developers in understanding how different parts of the system interact and how to implement or modify them effectively.
*Example:* Without proper documentation, understanding the integration of a new game into the platform could become a time-consuming and error-prone process.
***Question26:*** Is End-user Documentation available?
*Stakeholder*: Organizers & Participants
Score: 5
*Justification:* End-user documentation, such as tutorials and help guides, is important for user satisfaction and reducing support costs.
*Example:* Poorly documented features may lead to user frustration and increased support queries, impacting the platform's reputation.
***Question27:*** Is the documentation clear?
*Stakeholder*: Organizers & Participants
Score: 5
*Justification:* Clarity is essential for effective communication in documentation. Clear documentation enables users and developers to easily understand the information presented without ambiguity or confusion, enhancing usability and efficiency.
*Example:* Ambiguous documentation on how to submit scores for a competition could result in inconsistent data submissions, affecting the leaderboard's integrity.
***Question28:*** Is the documentation up to date?
*Stakeholder*: Organizers & Participants
Score: 5
*Justification:* Keeping documentation up to date is vital as outdated documentation can mislead developers and users, leading to the implementation of deprecated features or the use of older platform versions.
*Example:* If the documentation does not reflect recent changes to the competition rules, participants might follow outdated guidelines, leading to disqualification or confusion.

## C.9 Ethics debt

***Question29:*** Do you know the implementation rules, how to raise questions for clarification, and resolve conflicting interpretations of essentially contested concepts?
*Stakeholder*: Organizers & Participants
Score: 5
*Justification:* Knowing the implementation rules is crucial for ensuring that the competition is fair and that all participants have a clear understanding of what is expected. This knowledge helps maintain the integrity of the competition and prevents ethical breaches.
*Example:* In the RangL competition platform, clear implementation rules ensure that participants can optimize strategies within ethical boundaries.
***Question30:*** Do you know the consequences of non-compliance?
*Stakeholder*: Organizers & Participants
Score: 5
*Justification:* Awareness of the consequences of non-compliance is essential to deter unethical behavior and ensure adherence to the competition's rules. It also helps in maintaining a level playing field for all participants.
*Example:* In gaming competitions, non-compliance with rules can lead to disqualification or legal actions, as seen in cases where anti-competitive collusion resulted in fines and penalties. i.e. especially for Game Jams

## C.10 Infrastructure debt

***Question31:*** Are there mechanisms in place for automated monitoring and alerting of infrastructure performance metrics (e.g., CPU usage, memory utilization, network throughput)?
*Stakeholder*: Organizers
Score: 4
*Justification:* - Maintainability: Automated monitoring can reduce technical debt by making it easier to maintain the system.
- Future-proofing: Early detection of performance issues can prevent the accumulation of technical debt related to system degradation

*Example:* Implementing comprehensive monitoring from the start can avoid the need for costly refactoring of the monitoring system later on. Our competition's infrastructure includes an automated monitoring system that continuously tracks CPU and GPU usage, memory utilization, and network throughput. If any metric crosses a predefined threshold, such as CPU usage exceeding 90%, the system triggers an alert to our technical team, prompting immediate investigation and intervention to prevent any performance degradation during the competition.
***Question32:*** Has provision been made in the infrastructure for sufficient computing resources available?
*Stakeholder*: Organizers & Participants
Score: 3
*Justification:* - Scalability: While important, over-provisioning resources can lead to unnecessary complexity and costs, contributing to technical debt.
- Cost Management: Balancing resources with actual needs can minimize expenses and reduce the risk of investing in technologies that may become obsolete.

*Example:* Investing in scalable cloud services can prevent over-commitment to a particular infrastructure setup, reducing long-term technical debt.
***Question33:*** Have you developed a robust data pipeline for easy experimentation with AI algorithms?
*Stakeholder*: Organizers
Score: 5
*Justification:* Ensuring the competition platform provides necessary infrastructure for efficient experimentation is crucial. A robust data pipeline allows participants to quickly iterate, test various algorithms, and refine their models, enhancing the quality of submissions and boosting platform competitiveness. Effective infrastructure management is key to the success of AI-based competitions.
*Example:* Suppose a participant wants to explore different machine learning models for a computer vision task in an image recognition competition. Having a robust data pipeline allows them to effortlessly preprocess large datasets, train multiple models with different architectures and hyperparameters, and evaluate their performance, facilitating rapid experimentation and optimization.
***Question34:*** Have you automated pipelines for model training, deployment, and integration?
*Stakeholder*: Organizers & Participants
Score: 4
*Justification:* Automated pipelines are essential for managing AI models on the competition platform. They streamline processes for both organizers and participants by reducing manual effort, minimizing errors, and ensuring consistent model deployment and integration. This enhances the platform's scalability and efficiency, enabling seamless management of numerous submissions and models.
*Example:* Imagine an organizer hosting a competition where participants need to deploy their trained models to make predictions on new data. With automated pipelines in place, participants can simply upload their model artifacts to the platform, and the system automatically handles the deployment process, integrating the models into the competition framework for evaluation without manual intervention. This streamlines the submission process and accelerates the deployment of new models on the platform.

## C.11 Model debt

***Question35:*** Are you detecting direct feedback loops or hidden feedback loops?
*Stakeholder*: Organizers & Participants
Score: 4
*Justification:* Feedback loops can significantly distort the learning process, leading to suboptimal or even harmful behaviors in the model. Detecting them early is essential to prevent the amplification of biases and errors.
*Example:* In a game where an AI agent is trained to collect rewards, a direct feedback loop might occur if the agent learns to manipulate the game environment to generate rewards without performing the intended task. A hidden feedback loop could arise if the agent's actions inadvertently change the environment in a way that affects the agent's future state evaluations, leading to unintended strategies.
***Question36:*** Is model quality validated before serving?
*Stakeholder*: Participants
Score: 5
*Justification:* Validation ensures that the model performs as expected on unseen data and in real-world scenarios. It's a safeguard against deploying models that might perform well in training but fail in practice.
*Example:* Before deploying a model trained to play chess, it should be validated against a diverse set of opponents and scenarios to ensure it doesn't just exploit patterns seen during training but can generalize its strategy to new games.
***Question37:*** Does the model allow debugging by observing the step-by-step computation of training or inference on a single example?
*Stakeholder*: Participants
Score: 3
*Justification:* The ability to debug a model at a granular level is important for understanding and improving its decision-making process. However, it might be less critical than the other two questions if the model is performing well overall.
*Example:* If an AI agent makes an unexpected move in a game, being able to step through the computation process can help identify why that decision was made, whether it was due to a flaw in the model or an unforeseen aspect of the game environment.

## C.12 People – Social debt

***Question38:*** Is there a system in place to ensure project continuity through team member overlap and retention of the original development team's knowledge?
*Stakeholder*: Organizers
Score: 5
*Justification:* Ensuring project continuity through team member overlap and knowledge retention is paramount. This prevents the loss of expertise and maintains the quality and integrity of the platform.
*Example:* If a key developer leaves, their replacement can quickly get up to speed if there's comprehensive documentation and a system for knowledge transfer.
***Question39:*** Has a project support community been created?
*Stakeholder*: Organizers
Score: 3
*Justification:* Having a support community is still highly important. It fosters collaboration, user engagement, and can lead to community-driven development and troubleshooting.
*Example:* A forum where users can discuss strategies, report bugs, and suggest features can greatly reduce the burden on the core development team and help prioritize tasks based on user feedback.

## C.13 Process debt

***Question40:*** Have you correctly described your data handling procedures?
*Stakeholder*: Organizers
Score: 4
*Justification:* Data handling procedures are critical as they directly impact the quality of training data and subsequently affect the performance of the RL agent. Understanding how data is collected, preprocessed, and fed into the RL model is essential for ensuring the agent's effectiveness.
*Example:* Ensuring that the data collection process is unbiased and that the replay buffer is managed effectively to prevent overfitting or underutilization of data.
***Question41:*** Have you correctly described your model development processes?
*Stakeholder*: Organizers
Score: 4
*Justification:* Model development processes outline how the RL algorithm is designed and trained. This includes aspects such as the choice of algorithm, network architecture, hyperparameters, and training methodology. Understanding these processes is fundamental for reproducibility and ensuring the model's reliability and performance.
*Example:* Documenting the transition from a model-free to a model-based approach, detailing the algorithms used, and the rationale behind choosing specific hyperparameters.
***Question42:*** Have you correctly described the deployment processes of your model?
*Stakeholder*: Organizers
Score: 4
*Justification:* Deployment processes are crucial as they determine how the trained RL model is integrated into the game environment for real-world interaction. Understanding deployment procedures ensures smooth transition from development to production, minimizing potential errors and ensuring the model functions as intended in the game environment.

*Example:* Describing the process of integrating the trained RL model into the live game environment, including any safety checks and performance monitoring systems.

## C.14 Requirement debt

***Question43:*** Have you thoroughly defined the objectives, scope, stakeholder needs, expectations, decision goals, and insights of the AI system to ensure alignment with business objectives and user expectations?
*Stakeholder*: Organizers
Score: 5
*Justification:* Clearly defining the objectives, scope, stakeholder needs, and expectations is fundamental to the success of any AI system. If these aspects are not well-defined, it can lead to requirement debt, where the system does not meet the necessary criteria for success due to ambiguous or incomplete requirements.
*Example:* If the competition platform's goal is not aligned with the stakeholders' expectations, it may result in a system that is technically sound but fails to engage users or meet business objectives.
***Question44:*** Have you thoroughly addressed the technical aspects of the AI system, including the selection of appropriate AI techniques, algorithms, and models to achieve desired functionality and performance, as well as specifying quality attributes, trade-offs, metrics, and indicators to measure and evaluate system performance effectively?
*Stakeholder*: Organizers
Score: 5
*Justification:* Neglecting technical aspects such as AI techniques and models can result in a system that doesn't meet performance expectations, leading to requirement debt. Technical requirements are essential for system functionality.
*Example:* Choosing an overly complex model for a simple game could result in longer training times and difficulty in interpreting the model's decisions, making it harder to debug and improve.
***Question45:*** Have you monitored and retrained the AI system with new data as needed?
*Stakeholder*: Organizers & Participants
Score: 4
*Justification:* Continuously monitoring and retraining the AI system with new data is important for maintaining its relevance and performance, which can prevent the accumulation of technical debt due to model staleness or performance degradation.
*Example:* In a game competition platform, if new strategies or game updates are introduced, the AI must be retrained to understand these changes. Failing to do so can result in a system that performs poorly and requires significant refactoring later on.

## C.15 Self-Admitted Technical Debt (SATD)

***Question46:*** Have you addressed readability concerns in the RL model's code, such as poorly named variables, to improve maintainability?
*Stakeholder*: Participants
Score: 4
*Justification:* Readability of code improves maintainability, which is crucial for long-term development and debugging. However, this might not directly impact the RL model's performance.
*Example:* Renaming variables from var1 to player_score to clarify their purpose.
***Question47:*** Have you tackled code duplication in feature extraction code, ensuring efficiency and maintainability?
*Stakeholder*: Participants
Score: 4
*Justification:* Code duplication in feature extraction can affect both efficiency and maintainability, which are important aspects in RL applications.
*Example:* Refactoring two similar functions for extracting player statistics into a single reusable function.
***Question48:*** Have you minimized direct editing of ML model weights and layers to maintain model integrity and stability?
*Stakeholder*: Participants
Score: 3
*Justification:* Direct editing of model weights can lead to instability and unexpected behavior, making it crucial to minimize such actions.
*Example:* Using version control to track changes in model architecture instead of manual tweaks.
***Question49:*** Have you identified and addressed dependencies on external libraries or components that may hinder future changes?
*Stakeholder*: Participants
Score: 4
*Justification:* Dependencies on external libraries can impact the flexibility and maintainability of the system, thus requiring attention.
*Example:* Documenting and containerizing the current environment to ensure reproducibility.
***Question50:*** Have you identified and removed unnecessary or deprecated model code to streamline the system?
*Stakeholder*: Participants
Score: 3
*Justification:* Removing unnecessary or deprecated code improves maintainability and potentially performance by reducing complexity.
*Example:* Deleting legacy features that have been replaced by more efficient algorithms.
***Question51:*** Have you standardized data representation within the source code and storage mechanisms?
*Stakeholder*: Organizers & Participants
Score: 4
*Justification:* Standardizing data representation improves consistency and maintainability, which are important in RL development.
*Example:* Ensuring all player data is stored in a uniform structure across the platform.
***Question52:*** Have you established clear boundaries between subsystems to facilitate maintenance and changes?
*Stakeholder*: Organizers
Score: 4

*Justification:* Clear boundaries between subsystems improve maintainability and facilitate changes, which are important in complex RL systems.
*Example:* Defining separate modules for the matchmaking system and the scoring system.
***Question53:*** Have you designed the model code with maintenance and future modifications in mind?
*Stakeholder*: Participants
Score: 3
*Justification:* Designing the model code with maintenance and future modifications in mind is crucial for the long-term success of the RL system.
*Example:* Building a flexible architecture that allows for the easy addition of new game modes.
***Question54:*** Have you considered potential performance impacts of changes to the ML workflow?
*Stakeholder*: Participants
Score: 3
*Justification:* Considering performance impacts is crucial to ensure that changes to the ML workflow do not degrade system performance.
*Example:* Monitoring the impact of a new feature extraction method on the overall system latency.
***Question55:*** Have you implemented measures to ensure robustness of the RL model against variations in data quality?
*Stakeholder*: Participants
Score: 4
*Justification:* Ensuring robustness against variations in data quality is essential for the reliability of the RL model.
*Example:* Implementing data validation checks to detect and handle anomalous input data.

## C.16 Test debt

***Question56:*** Have all hyperparameters been correctly tuned, validated, and ensured to be optimal for performance in the game environment?
*Stakeholder*: Participants
Score: 5
*Justification:* Tuning hyperparameters directly affects the performance of the RL agent in the game environment. It's crucial for optimizing its behavior. Optimal hyperparameters are vital for the RL agent to effectively learn and adapt to the game dynamics. Correct hyperparameter tuning is essential in RL game applications to optimize the model's performance and enhance the gaming experience, ensuring efficient learning and effective decision-making in-game scenarios.
*Example:* Tuning the learning rate can significantly affect the convergence speed and stability of the training process.
***Question57:*** Has reproducibility of agent training and environment dynamics been tested to ensure consistency?
*Stakeholder*: Organizers & Participants
Score: 4
*Justification:* Reproducibility ensures consistency and reliability in training the agent, which is crucial for fair gameplay.
*Example:* The use of fixed random seeds to ensure that results can be replicated across different runs.
***Question58:*** Is there a fully automated test regularly running to validate the entire pipeline, ensuring data and code move through each stage successfully and resulting in a well-performing model?
*Stakeholder*: Organizers
Score: 5
*Justification:* Having a fully automated test ensures the integrity of the entire pipeline, which is essential for maintaining the quality of the game.
*Example:* Automated regression tests can catch issues early before they affect the model's performance.
***Question59:*** Do the data invariants hold for the inputs in the game environment?
*Stakeholder*: Organizers & Participants
Score: 3
*Justification:* Data invariants are important for maintaining the integrity of the game environment and ensuring fair gameplay.
*Example:* Checking that the positions of players in a sports game do not suddenly jump to unrealistic values.
***Question60:*** Are there mechanisms in place to ensure that training and serving are not skewed in the game?
*Stakeholder*: Organizers & Participants
Score: 4
*Justification:* Skewed training and serving can lead to unfair advantages or disadvantages for players.
*Example:* Using the same feature engineering pipeline for both training and serving can help avoid discrepancies.
***Question61:*** Are the models numerically stable for effective gameplay?
*Stakeholder*: Organizers & Participants
Score: 5
*Justification:* Numerically stable models ensure reliable and consistent behavior during gameplay.
*Example:* The use of gradient clipping in training to prevent exploding gradients.
***Question62:*** Has the prediction quality of the game not regressed over time?
*Stakeholder*: Organizers & Participants
Score: 5
*Justification:* Prediction quality directly impacts the agent's decisions and ultimately the gameplay experience. Therefore, maintaining its quality is crucial.
*Example:* Tracking the accuracy of the model's predictions against a validation set over multiple seasons of a game.

## C.17 Versioning debt

***Question63:*** Have you installed a proper version control system for model, training and test data?
*Stakeholder*: Organizers
Score: 5
*Justification:* Version control is crucial in any software development project, including RL game applications, to keep track of changes, revert to previous states if necessary, and collaborate effectively. Without version control, it can be challenging to manage changes, leading to potential errors and difficulties in reproducing results.
*Example:* If a new model version causes a regression in performance, a version control system would allow developers to quickly revert to a previous, stable version.
***Question64:*** Have you used the appropriate policy for marking the versions of your software components?
*Stakeholder*: Organizers
Score: 3
*Justification:* Proper versioning allows for clear communication about changes and updates to software components. It helps users understand the significance of updates (major, minor, or patch) and ensures compatibility across different versions. While essential, it may not directly impact the RL game's functionality as much as version control itself.
*Example:* If a major update is released without proper version marking, it could break compatibility with existing systems that rely on the platform, leading to significant technical debt.
***Question65:*** Do you maintain a consistent data structure for game state representation throughout iterations, ensuring compatibility between different versions of the RL game?
*Stakeholder*: Organizers & Participants
Score: 4
*Justification:* Ensuring consistency in the data structure for representing the game state is crucial in RL game development. Changes in the data structure could affect the performance of RL algorithms, training stability, and overall gameplay experience. Keeping track of these changes and ensuring compatibility between different versions of the game state representation can help maintain the integrity and effectiveness of the RL algorithms employed in the game.
*Example:* If the game state representation changes without maintaining consistency, models trained on previous versions of the data may become obsolete, requiring retraining or adaptation.

## C.18 Accessibility debt

***Question66:*** Have you conducted usability testing to identify and address potential barriers in the platform setup process?
*Stakeholder*: Organizers
Score: 5
*Justification:* Conducting usability testing is critical to uncover and address accessibility issues that can impede participants' engagement. By identifying these barriers early, organizers can make necessary adjustments to enhance the platform's usability and ensure a smooth user experience.
*Example:* If usability testing reveals that participants struggle with the initial setup due to unclear instructions, organizers can simplify the process and provide more intuitive guidance, thereby reducing Accessibility Debt and improving participant retention.
***Question67:*** Have you integrated adaptive user interfaces that personalize the setup experience based on participants' skill levels and preferences?
*Stakeholder*: Organizers
Score: 4
*Justification:* Integrating adaptive user interfaces can significantly enhance usability by tailoring the setup process to individual participants' needs. This personalized approach can reduce frustration and improve engagement by providing relevant guidance and support based on users' varying levels of expertise.
*Example:* If a beginner is guided through a simplified setup process with additional tips and resources, while an advanced user is offered a streamlined version, both users are more likely to have a positive experience and continue their participation without unnecessary obstacles.
***Question68:*** Have you implemented feedback mechanisms for participants to report accessibility issues and suggest improvements?
*Stakeholder*: Organizers
Score: 3
*Justification:* Implementing feedback mechanisms allows organizers to continuously improve the platform based on user input. By actively listening to participants' concerns and suggestions, organizers can promptly address accessibility issues and enhance the overall user experience.
*Example:* If participants can easily report setup difficulties or suggest enhancements, organizers can quickly implement changes, thereby minimizing Accessibility Debt and fostering a more user-friendly environment.